\documentclass[10pt,letterpaper]{article}
\usepackage[OE]{express}
\usepackage{graphicx}
\usepackage{cite}
\usepackage{color}
\usepackage{amsmath}
\usepackage{here}
\newcommand{\be}{\begin{equation}}
\newcommand{\ee}{\end{equation}}
\newcommand{\bea}{\begin{eqnarray}}
\newcommand{\eea}{\end{eqnarray}}

\begin{document}

\title{Generalized Kerker effects in nanophotonics and meta-optics}

\author{Wei Liu,$^{1,*}$ and Yuri S. Kivshar$^{2,3,4}$}
\address{
$^1$College of Optoelectronic Science and Engineering, National University of Defense
Technology, Changsha, Hunan 410073, P. R. China\\
$^2$Nonlinear Physics Centre, Australian National University, ACT 2601, Australia\\
$^3$ITMO University, St. Petersburg 197101, Russia\\
$^4$Email:yuri.kivshar@anu.edu.au}
\email{\authormark{*}wei.liu.pku@gmail.com}
\begin{abstract}
The original Kerker effect was introduced for a hypothetical magnetic sphere, and initially it did not attract much attention due to a lack of magnetic materials required. Rejuvenated by the recent explosive development of the field of metamaterials and especially its core concept of optically-induced artificial magnetism, the Kerker effect has gained an unprecedented impetus and rapidly pervaded different branches of nanophotonics. At the same time, the concept behind the effect itself has also been significantly  expanded and generalized. Here we review the physics and various manifestations of the generalized Kerker effects, including the progress in the emerging field of meta-optics that focuses on interferences of electromagnetic multipoles of different orders and origins. We discuss not only the scattering by individual particles and particle clusters, but also the manipulation of reflection, transmission, diffraction, and absorption for metalattices and metasurfaces, revealing how various optical phenomena observed recently are all ubiquitously related to the Kerker's concept.
\end{abstract}



\section{Introduction}

The history of the Kerker effect goes back to 1983 when M. Kerker~\textit{et al.} studied  scattering of a magnetic sphere with magnetic permeability $\mu\neq1$~\cite{Kerker1983_JOSA}. One of the significant findings of this study was the observation that, when the electric permittivity  satisfies the condition $\epsilon=\mu$, the backscattering by the magnetic sphere can be totally eliminated. This effect can be attributed to the fact that two sets of the Mie scattering coefficients coincide for every expansion order~~\cite{Kerker1983_JOSA}. If we take into account the correspondence between the Mie coefficients and electromagnetic multipoles~\cite{hulst_light_1957,jackson1962classical,Bohren1983_book}, a direct conclusion from ~\cite{Kerker1983_JOSA} is a particle that supports overlapped in-phase electric and magnetic multipoles of the same order and magnitude would not scatter backwards. The simplest case of this effect is overlapped electric and magnetic dipole resonances, which corresponds exactly to the so-called Huygens' source employed for antenna engineering~\cite{Jin2010_IEEE}. Unfortunately, the discovery made by M. Kerker~\textit{et al.} had gradually slipped into oblivion, as there exist almost no materials that are naturally magnetic, especially in the spectral regimes of higher frequencies.

Revitalized by the concept of optically-induced artificial magnetism~\cite{Pendry1999_ITMT}, and especially the demonstration with simple high-index dielectric particles~\cite{Zhao2009_materialtoday,Kuznetsov2012_SciRep,Evlyukhin2012_NL,Liu2014_CPB}, the situation has been dramatically changed, with the original work by M. Kerker~\textit{et al.} attracting surging attention and interest in various branches of nanophotonics. This has breathed a new life into the original proposal for magnetic spheres, which has then been greatly extended to nonmagnetic single particles of other shapes and finite particle clusters, to control scattering not only at the originally discussed forward and backward directions, but also along all other possible scattering angles. Consequently, a new branch of nanophotonics, termed as  meta-optics, has been incubated focusing on optical-wave manipulations replying on resonant multipolar excitations and interferences involving optically-induced magnetic responses.  This field is rapidly developing, and it finds various applications not only with single particles and finite particle clusters, but also with other extended periodic or aperiodic structures ~\cite{Zhao2009_materialtoday,Liu2014_CPB,CHEN_Rep.Prog.Phys._review_2016,KUZNETSOV_Science_optically_2016,LIU_ArXivPrepr.ArXiv160901099_multipolar_2016,STAUDE_NatPhoton_metamaterialinspired_2017,KIVSHAR_Opt.PhotonicsNews_metaoptics_2017,KRUK2017ACSPhotonics,YANG2017PhysicsReports,DING_Rep.Prog.Phys._gradient_2017}.
Compared to the original Kerker's proposal, the term of \textit{generalized Kerker effects} can be justified mainly by, but not limited to, the following extensions: (i) The exciting source can be other more sophisticated structured electromagnetic waves or even electron beams. (ii) The scattering body can be isolated particles of arbitrary shapes, particle clusters, and periodic (quasi-periodic) particle lattices. (iii) Angular scattering pattern shaping can be also achieved in other scattering angles beyond the conventional forward and backward directions only. (iv) The interfering multipoles have been extended from mainly dipoles to higher-order multipoles, and interferences between pure electric resonances can be also applied for scattering shaping (in generalized cases, magnetic responses are not really essential anymore; see Sec. 2 for more details). This paper aims to discuss, for the first time to our knowledge, a broad variety of problems driven by the generalized Kerker effects, and we reveal how those effects play ubiquitous and significant roles in various optical phenomena and related advanced applications.

To realize these ideas, first we discuss the far-field radiation phase symmetries of electromagnetic multipoles and demonstrate how different combinations of them can provide a broader theoretical framework for scattering pattern shaping. Then,  we examine the generalized Kerker effects in individual particles and finite particle clusters, discussing the manipulations of both differential and total scattering cross sections. We then move to the most interesting generalizations in periodic structures, that we have termed here as {\em metalattices} underlying their differences from metasurfaces and metagratings. We demonstrate how the generalized Kerker effects are directly related to various functionalities in meta-optics, including perfect transmission, perfect reflection, higher-order diffraction management, and perfect absorption. We conclude the review with the discussions of perspectives, interdisciplinary connections and possible broader applications.

\section{Phase-symmetry analysis for electromagnetic multipoles}

\begin{figure*}
\centerline{\includegraphics[width=12cm]{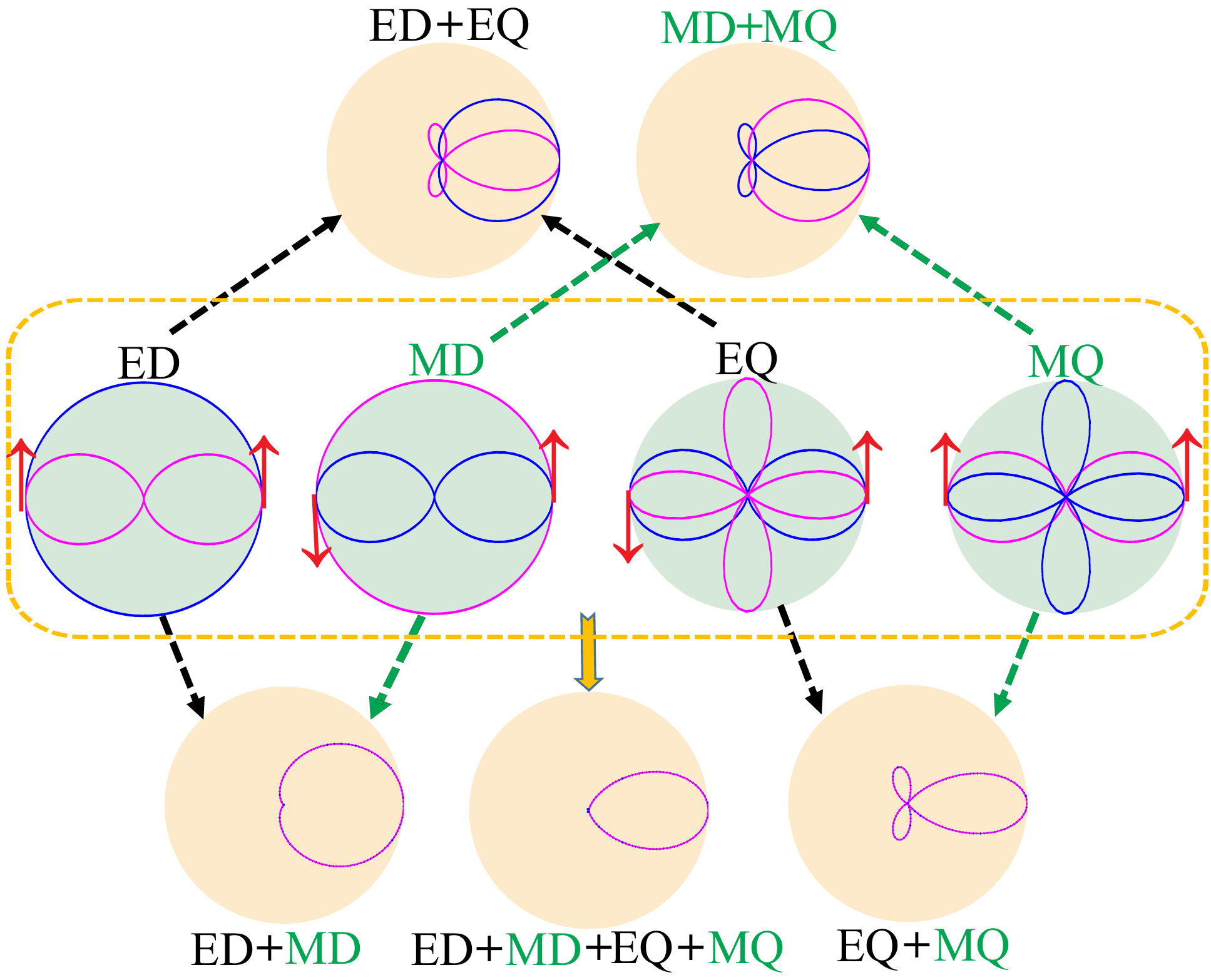}}\caption{\small Phase-symmetry analysis for electromagnetic multipoles up to quadrupoles (middle row; arrows indicate radiated electric fields in the forward and backward directions, with upward and downward arrows corresponding to out-of-phase and in-phase fields with respect to the incident electric field, respectively) and different overlapping scenarios to suppress the backward scattering (upper and lower rows). The incident plane wave is assumed to prorogate from the left with in-plane electric field. All the multipoles shown are resonantly excited and of the same magnitude in terms of backward scattering. For both individual and overlapped multipoles, only the in-plane (purple curves) and out-of-plane (blue curves) scattering patterns are shown for clarity.  The scattering patterns are azimuthally symmetric (in-plane and out-of-plane scattering patterns are identical) for overlapped electric and magnetic multipoles of the same order (lower row).}
\label{fig1}
\end{figure*}

 For the physics of electromagnetic multipoles, we would usually consider their two representative features: the corresponding charge-current distributions and far-field radiation patterns. This approach brings us to \textit{a widely spread misconception} about the hierarchy of the electromagnetic multipoles: it is usually believed that the electric dipole (ED) is the fundamental mode that dominates over other multipoles, such as the magnetic dipole (MD) mode,\textit{ etc.}.  This taken-for-granted hierarchy comes from the Taylor expansion for the vector potential of a specific charge-current distribution, which can be physically justified only in the large-wavelength approximation when the source dimension is much smaller than the effective wavelength~\cite{jackson1962classical,Radescu2002_PRE,liu_efficient_2015,alaee_electromagnetic_2018}. When this precondition is not satisfied, the physical validity of the Taylor expansion and thus the multipole hierarchy becomes not well-grounded, indicating that the ED mode would not necessarily be the dominant and strongest mode anymore. A noticeable example of this is the recent demonstration of optically-induced magnetism with high-index (thus with relatively smaller effective wavelength) dielectric particles, where the MD modes can be made comparable or even stronger than ED modes in some spectral regimes in terms of the total scattered power~\cite{Kuznetsov2012_SciRep,Evlyukhin2012_NL}. Moreover, it is revealed that simple particles (not only dielectric but also metallic or hybrid) can provide a full family of electromagnetic multipoles beyond dipoles, such as electric quadrupole (EQ) and magnetic quadrupole (MQ) modes, which renders tremendous opportunities for the observation of different types of the generalized Kerker effects relying on interferences between multipoles of different natures (electric or magnetic) and orders~\cite{Zhao2009_materialtoday,Liu2014_CPB,KUZNETSOV_Science_optically_2016,LIU_ArXivPrepr.ArXiv160901099_multipolar_2016,STAUDE_NatPhoton_metamaterialinspired_2017}.

Another significant feature of electromagnetic multipoles that has been largely neglected in the past is associated with the phase symmetry of their far-field radiations, though the symmetries become essential for the study of multipolar interferences. A systematic study reveals that multipoles show very different phase symmetries in the forward and backward directions with respect to the incident wave~\cite{liu_ultradirectional_2014}. Basically, the ED radiation shows even parity while the MD radiation shows odd parity, for which the radiated electric fields in the forward and backward directions are in and out of phase, respectively~\cite{jackson1962classical,hulst_quantifying_2012}.  We notice here that the parities are discussed in terms of electric fields, while for magnetic fields the parities would be reversed due to the electromagnetic duality, as is also the case for other multipoles. The phase symmetries of higher-order multipoles can be simply deduced from the following simple rule: the scattering parities are opposite, for multipoles of the same nature and adjacent orders (\textit{e.g.}, even parity, for ED, while odd parity, for EQ), and for multipoles of the same order but different natures (\textit{e.g.}, odd for EQ while even for MQ).  Though not widely recognised, this simple rule manifests itself in many well-known mathematical expressions~\cite{LIU_Phys.Rev.Lett._generalized_2017}. For example, the backward scattering intensity for spherical particles with an incident plane wave is proportional to $|\sum\nolimits_{n = 1}^\infty  {(2n + 1} ){( - 1)^n}({b_n} - {a_n}){|^2}$, where $a_n$ and $b_n$ are Mie scattering coefficients of the order $n$~\cite{Kerker1983_JOSA,hulst_light_1957,jackson1962classical,Bohren1983_book}. It is clear that this expression is consistent with the above mentioned multipole phase parity rule:  the terms of $(-1)^n$  and (${b_n} - {a_n}$)  correspond to the opposite parities for the multipoles of the adjacent orders and of different natures, respectively.

In Fig.~\ref{fig1}, we show schematically the far-field radiation phase parities for resonantly excited multipoles up to quadrupoles (middle row), where we assume that the exciting plane wave propagates from the left, with an in-plane electric field pointing downwards. For each multipole, we show for clarity only the in-plane (purple curve) and out-of-plane (blue curve) scattering patterns. The radiated electric fields of all multipoles interfere destructively in the forward direction with respect to the incident wave (all upward pointing arrows), as is required by the optical theorem~\cite{hulst_light_1957,jackson1962classical,Bohren1983_book}. As for the backward radiated electric fields, the phase would be decided by the parity of each multipole. It is clear here that, to suppress the backward scattering, we can not only overlap ED and MD modes (as is the simplest case of Huygens' source~\cite{Jin2010_IEEE}), but also more generally overlap multipoles of opposite parities (multipoles of same nature but different orders, or multipoles of the same order but different natures; see upper and lower rows in Fig.~\ref{fig1}). Moreover, to overlap more electric and magnetic multipoles of higher orders would not only suppress the backward scattering (and other side scattering lobes; see \textit{e.g.} the case in the lower row of Fig.~\ref{fig1} with overlapped ED, MD, EQ and MQ modes), but also enhance the directionality of the forward scattering~\cite{liu_ultradirectional_2014,Liu2015_OE_Ultra}. The forward scattering lobe can be further collimated through employing engineered arrays of overlapped multipoles~\cite{Liu2012_ACSNANO,WANG_ACSNano_janus_2015,ziolkowski_using_2017}.

It is worth mentioning that in Fig.~\ref{fig1} we present only the radiation phase in the forward and backward directions. Moreover, we have confined our discussions to resonantly excited multipoles by the same incident wave (all multipoles are in-phase); the overlapped multipoles are of the same magnitude (in terms of backward scattering) and of opposite parities to suppress the overall backward scattering. Definitely, the phase analysis can be extended to other scattering angles, and we can generalize the interferences to multipoles with arbitrary relative phases and amplitudes (including the so-called ``second Kerker's condition" to suppress the forward scattering), or even to those of the same parity, to enhance the backward scattering~\cite{Kerker1983_JOSA,Alu2010_JN,Gomez-Medina2011_JN,Geffrin2012_NC,Fu2013_NC,liu_scattering_2013-1,ZAMBRANA-PUYALTO2013Opt.Lett.,liberal_superbackscattering_2015,PANIAGUA-DOMINGUEZ_NatCommun_generalized_2016-1,LIU_Phys.Rev.A_superscattering_2017,wiecha_strongly_2017,lee_reexamination_2017}. All those features constitute the main aspects of the generalized Kerker effects, which can render more flexibilities for the scattering manipulation than the initially proposed Kerker effect~\cite{Kerker1983_JOSA}.

\section{Generalized Kerker effects for individual particles}

\begin{figure*}
\centerline{\includegraphics[width=13cm]{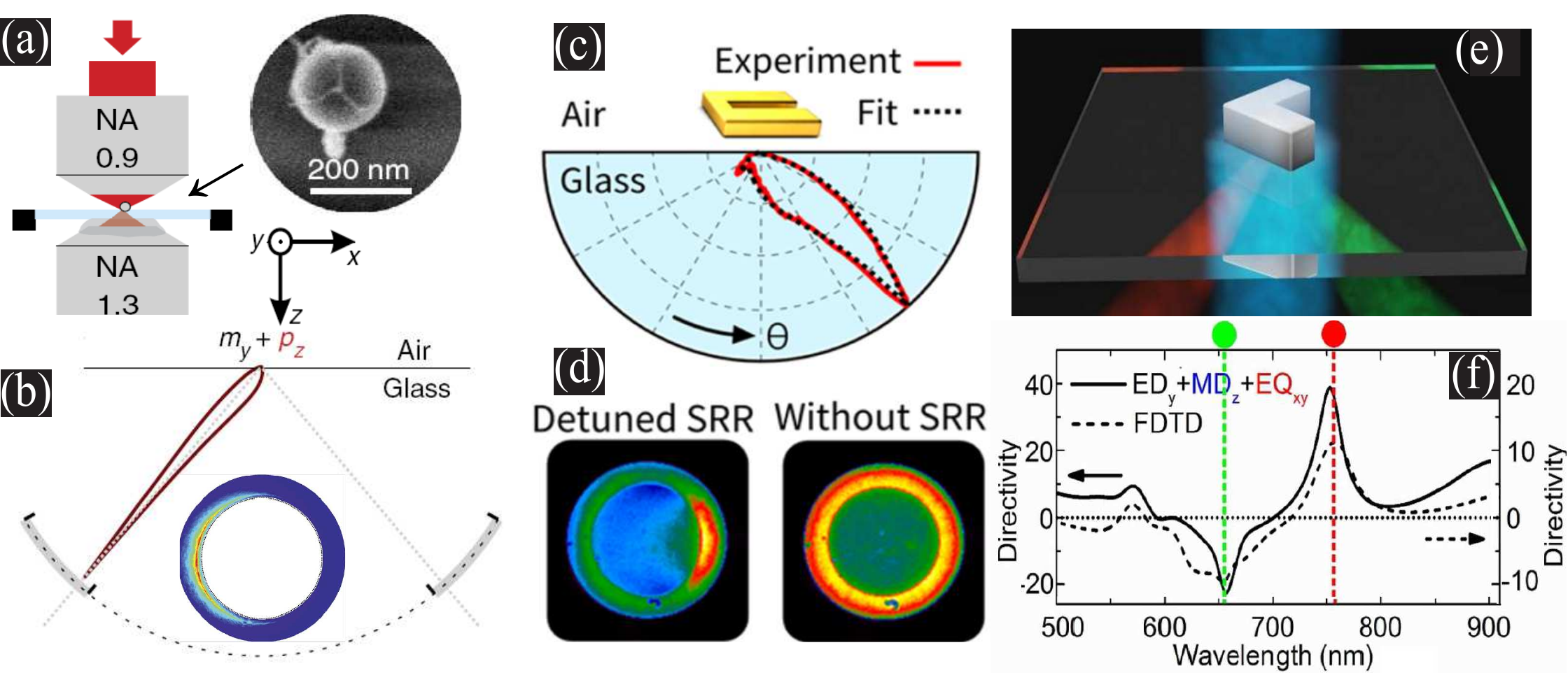}}\caption{\small Directional scattering by individual particles, induced by interference of: (a,b) ED and MD modes, and (c-f) ED, MD, and EQ modes. In (a,b), the ED and MD modes are excited in a Si nanoparticle by a focused radially polarized wave. In (c-f), ED, MD and EQ modes are excited by (c,d) quantum dots within a gold SRR, and (e,f) by an effective plane waves within the V-shaped silicon antenna.  Adapted from [43,53,54].}
\label{fig2}
\end{figure*}

As already mentioned, the original Kerker effect was introduced for a single magnetic sphere in free space with an incident plane wave~\cite{Kerker1983_JOSA}. It has been greatly generalized, not only to various nonmagnetic particles of different shapes~\cite{Zhao2009_materialtoday,Liu2014_CPB,KUZNETSOV_Science_optically_2016,LIU_ArXivPrepr.ArXiv160901099_multipolar_2016,STAUDE_NatPhoton_metamaterialinspired_2017},
but also to other engineered incident waves~\cite{GOUESBET_generalized_2011,WANG_Opt.Lett._optimized_2014,WOZNIAK_LaserPhotonicsRev._selective_2015,NEUGEBAUER_NatCommun_polarizationcontrolled_2016,WEI_Opt.Lett.OL_adding_2017,
XI_Phys.Rev.Lett._magnetic_2017,LIU_Opt.ExpressOE_optical_2017,MELIK-GAYKAZYAN_2017_ACSPhotonics_Selective}, with the effects of substrates also studied in detail for practical applications~\cite{Novotny2012_book,GROEP2013Opt.Express,PORS_Opt.Express_unidirectional_2015,SINEV2016Laser&PhotonicsReviews,HOLSTEEN2017Science}. Figure~\ref{fig2} includes three experimental studies on efficient beam steering based on interferences of multipoles excited with individual particles~\cite{NEUGEBAUER_NatCommun_polarizationcontrolled_2016,Hancu2013_NL,LI_NanoLett._alldielectric_2016}. Figures~\ref{fig2}(a) and \ref{fig2}(b) show the directional scattering of a silicon particle into the substrate, based on the sole interference of electric and magnetic dipoles. Compared to conventional studies into dipolar interferences with a plane wave incidence~\cite{Zhao2009_materialtoday,Liu2014_CPB,KUZNETSOV_Science_optically_2016,LIU_ArXivPrepr.ArXiv160901099_multipolar_2016,STAUDE_NatPhoton_metamaterialinspired_2017}, here the incoming wave is a tightly focused radially-polarized beam, with an adjustable focusing point relative to the particle position. This renders extra freedom, compared to a linearly-polarized plane wave, for the tuning of both the amplitudes and phases of the excited multipoles ~\cite{WOZNIAK_LaserPhotonicsRev._selective_2015,NEUGEBAUER_NatCommun_polarizationcontrolled_2016}, thus providing more flexibilities to induce highly-efficient directional scattering. The same effect can also be observed with more interfering multipoles of various orders (see, \textit{e.g.} Fig.~\ref{fig1}), and two examples are shown in Figs.~\ref{fig2}(c)-\ref{fig2}(f)~\cite{Hancu2013_NL,LI_NanoLett._alldielectric_2016}. For both cases, the efficient light routing in the substrate is induced by interferences involving mainly ED, MD, and EQ modes,  which are excited (by quantum dots) in the classical gold split-ring resonator [SRR, Figs.~\ref{fig2}(c) and \ref{fig2}(d)] or are excited (by effectively plane waves) in the all-dielectric V-shaped silicon antenna [Figs.~\ref{fig2}(e) and \ref{fig2}(f)], respectively. For the V-shaped silicon antenna, at different wavelengths, the amplitude and phase ratios between ED, MD and EQ modes would be different, and thus there will be different interference patterns, routing light of different colors to different directions [Figs.~\ref{fig2}(e) and \ref{fig2}(f)]. It is clear here, the fundamental point that really matters, is not directly what materials the particles are made of or what is the shape, but which multipoles are excited and how they interfere with one another. Similarly, as long as the required multipoles (of proper phases and magnitudes) can be effectively excited, the directional scattering can also be achieved with various other configurations~\cite{GOUESBET_generalized_2011,WANG_Opt.Lett._optimized_2014,WOZNIAK_LaserPhotonicsRev._selective_2015,WEI_Opt.Lett.OL_adding_2017,
XI_Phys.Rev.Lett._magnetic_2017,LIU_Opt.ExpressOE_optical_2017}, including electron beam excitations and electrical excitations ~\cite{COENEN_NatCommun_directional_2014,WANG2016Phys.Rev.Lett.,GURUNARAYANAN2017NanoLett.}.

In Fig.~\ref{fig2} we confine our discussions to interfering multipoles up to quadrupoles, and it is known that introducing more multipoles of higher orders will render extra flexibilities for scattering shaping and beam controls (see, \textit{e.g.} Fig.~\ref{fig1})~\cite{liu_ultradirectional_2014,Liu2015_OE_Ultra,Rolly2013_arxiv,Krasnok2013_arxiv,VERCRUYSSE_ACSNano_directional_2014,NARAGHI_Opt.Lett.OL_directional_2015}. The same principles can also be applied to complementary particles (such as holes and slits~\cite{DEABAJO_Rev.Mod.Phys._colloquium_2007,KIHM_Nat.Commun._bethehole_2011,GROSJEAN_NanoLett._diabolo_2011,COENEN_ACSNano_optical_2014,ROTENBERG_NatPhoton_mapping_2014,WANG2016Phys.Rev.Lett.,YAN2017LightSci.Appl.}), spoof plasmonic structures~\cite{HUIDOBRO_Phys.Rev.X_magnetic_2014,WU2018ArXiv180104040Phys.} and particles incorporating tunable materials~\cite{SAUTTER2015ACSNano,ALAEE_Opt.Lett._phasechange_2016,WUTTIG_NatPhoton_phasechange_2017}. The generalized Kerker effects with individual particles can be employed not only for scattering shaping and beam control, but also for further related applications, such as more sophisticated manipulations of quantum dot emissions~\cite{RUTCKAIA2017NanoLett.,BARANOV2017Laser&PhotonicsReviews}, optical polarizations~\cite{KRUK_ACSPhotonics_spinpolarized_2014,GARCIA-ETXARRI2017ACSPhotonics}, and optical forces~\cite{Gomez-Medina2011_JN,NOVITSKY_Phys.Rev.Lett._materialindependent_2012,WANG_Opt.Lett._optimized_2014,LIU_Opt.ExpressOE_optical_2017,GAO_LightSci.Appl._optical_2017}.

We emphasize here that both the phase symmetry analysis and the interference patterns shown in Fig.~\ref{fig1} are valid only in the far field region, where the radiated fields are treated as transverse waves (no electric or magnetic components along the propagation direction), with all the evanescent components neglected~\cite{jackson1962classical}. Similarly, the different scenarios of directional radiations into the substrates shown in Fig.~\ref{fig2} are also attributed to the far-field interference effects between multipoles of different sorts and orders. In a sharp contrast, in the near field  both the field components (including both prorogating and evanescent ones) for each multipole and the interferences between different multipoles can be rather complicated. The near-field phase symmetry analysis for resonant excitations and interferences of various multipoles are vitally important for  nanoscale light-matter interactions, especially at the interfaces~\cite{DEABAJO_Rev.Mod.Phys._colloquium_2007,ROTENBERG_Phys.Rev.Lett._plasmon_2012,ROTENBERG_NatPhoton_mapping_2014,LEE_Phys.Rev.Lett._role_2012,RODRIGUEZ-FORTUNO_Science_nearfield_2013,EVLYUKHIN_Phys.Rev.B_resonant_2015,SINEV_Laser&PhotonicsReviews_chirality_2017,PICARDI_Phys.Rev.B_unidirectional_2017,
PICARDI_ArXiv170802494Phys._janus_2017}. Unfortunately, for higher-order multipoles especially, it has not been investigated in a comprehensive way to provide a clear and intuitive picture that is as exhaustive as what has been obtained in the far field (see, \textit{e.g.} Fig.~\ref{fig1}).

\begin{figure*}
\centerline{\includegraphics[width=13cm]{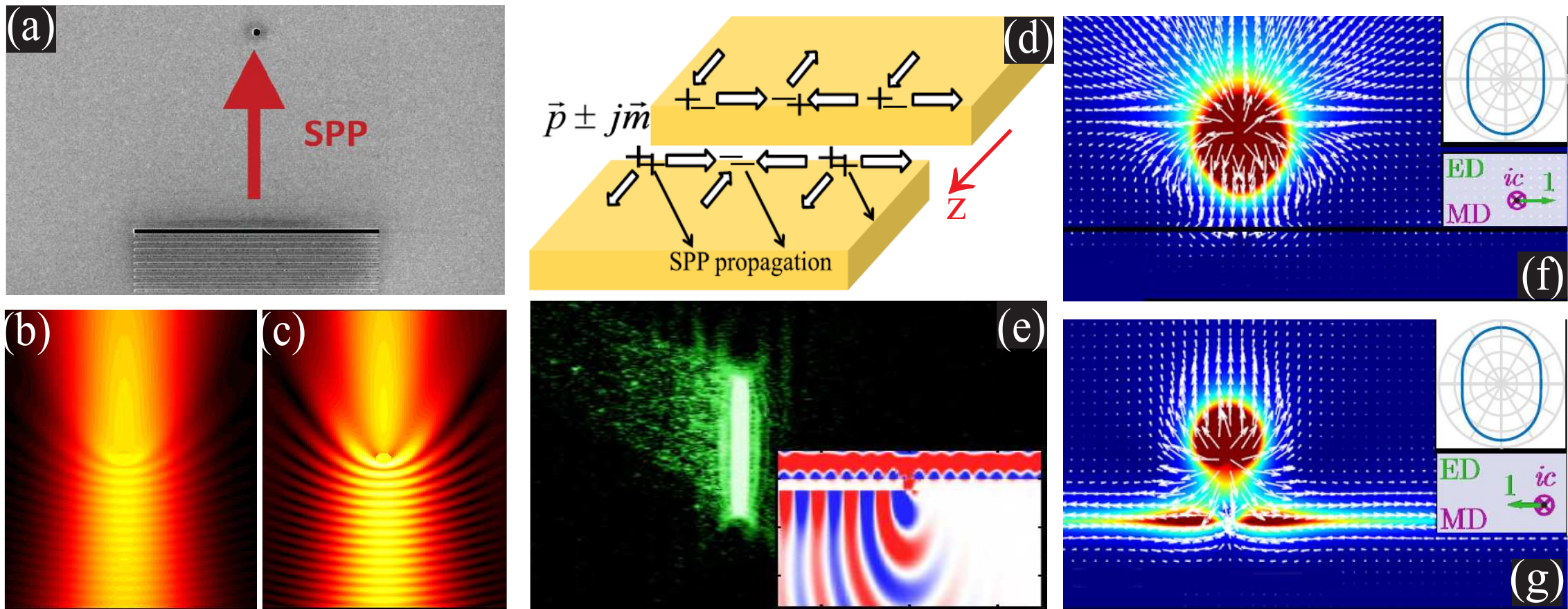}}\caption{\small Interference of an incident surface plasmon with the fields scattered by a hole in a gold film shown in (a), where both the ED and MD modes can be effective excited. The interference patterns are shown in (b) and (c), where the sole contribution of ED, or both the contributions from ED and MD are considered, respectively. Unidirectional surface plasmon wave excitation [shown in (e)] can be obtained with a metal slit [shown in (d)] excited by an obliquely-incident circularly-polarized wave. The directionality comes from the interference of a pair of parallel ED and MD [both along $z$ direction, shown in (d)] with an absolute $\pi/2$ phase shift that is induced by the circular polarization of the incident wave. Effective switching of the excitations of surface modes (or waveguide modes) through the interferences of a pair of orthogonal ED and MD, as is shown in (f,g). Through tuning the phase difference between the two dipoles from $\pi/2$ to $-\pi/2$, the mode excitations can be switched on or off, despite the fact that the far-field radiation patterns of such a dipole pair are identical for the two opposite phase differences [see the insets of (f,g)].Adapted from [79,80,85].}
\label{fig3}
\end{figure*}

To exemplify the significant role of the near-field multipolar interference, in Fig.~\ref{fig3} we show three cases involving excitation and propagation of surface waves~\cite{ROTENBERG_Phys.Rev.Lett._plasmon_2012,LEE_Phys.Rev.Lett._role_2012,PICARDI_ArXiv170802494Phys._janus_2017}. A single hole in a gold film shown in Fig.~\ref{fig3}(a) can support both ED and MD modes when excited with a surface plasmon wave~\cite{ROTENBERG_Phys.Rev.Lett._plasmon_2012}. Figures~\ref{fig3}(b) and \ref{fig3}(c) show interference patterns of the incident and scattered waves by a hole, considering respectively only the contribution of ED mode [Fig.~\ref{fig3}(b)], or both ED and MD modes [Fig.~\ref{fig3}(c)].  It is clear that the interference between the ED and MD modes is vitally important, especially for the interference fringes at larger distances from the beam central line. Figures~\ref{fig3}(d) and \ref{fig3}(e) show the unidirectional excitations of surface plasmons with a metal slit excited by an obliquely-incident circularly-polarized wave~\cite{LEE_Phys.Rev.Lett._role_2012}. It is observed that the directionality comes from the near-field interference of a pair of parallel ED and MD modes [both of them are along $z$ direction shown in Fig.~\ref{fig3}(d), with an absolute $\pi/2$ phase shift induced by the incident wave], which cannot be possibly obtained through the far-field interference of such a dipole pair configuration. Figures~\ref{fig3}(f) and \ref{fig3}(g) show the effective switching for the excitations of surface waves (or other waveguide modes), which has been made possible by the interference of a pair of ED and MD modes, with also an absolute $\pi/2$ phase shift~\cite{PICARDI_ArXiv170802494Phys._janus_2017}. Compared to what is shown in Figs.~\ref{fig3}(d) and \ref{fig3}(e), now the ED and MD modes are orthogonal rather than parallel to each other. Such a ED-MD pair is also different from the conventional Huygens' source~\cite{Jin2010_IEEE}, for which the two consisting dipoles are in phase. As a result, the far-field radiation patterns [see the insets in Figs.~\ref{fig3}(f) and \ref{fig3}(g), which are identical for the phase shift of both $\pi/2$ and $-\pi/2$ ] are also contrastingly different from the highly directional pattern of a Huygens' source (Fig.~\ref{fig1}).  Despite the identical far-field patterns, the near-field coupling of such an ED-MD pair to the surface modes are contrastingly different for opposite phase shifts [Figs.~\ref{fig3}(f) and \ref{fig3}(g)], confirming the fundamental differences between near-field and far-field multipolar interference effects.

We note here that the results summarized in Figs.~\ref{fig3}(d)-\ref{fig3}(g) are also related to or even beyond the physics of spin-orbit interactions of light~\cite{BLIOKH_Nat.Photonics_spinorbit_2015,VANMECHELEN_Optica_universal_2016,PICARDI_Phys.Rev.B_unidirectional_2017}. Moreover, it is natural to expect that, similar to what is shown in Fig.~\ref{fig1}, the study of near-field interference between dipoles can be extended to other higher-order multipoles.

\section{Generalized Kerker effects for particle clusters}

The analysis of the generalized Kerker effects applied to clusters of particles is less straightforward than that for individual particles, mainly due to inter-particle interactions. The interactions can make the multipolar efficiencies of each particle in the cluster totally different from the case when the particle is isolated. The general theory that can be employed to treat the particle clusters is the multiple scattering theory~\cite{QUINTEN_optical_2010}, through which the multipolar efficiencies within each particle of the cluster can be obtained. Then the scattering properties of the whole cluster can be obtained through a linear combination of the contributions from all the multipoles excited within all the particles of the cluster, taken into consideration also the phase lags among the particles. One of the simplest versions of multiple scattering theory is the coupled dipole theory that involves not only ED modes but also MD modes~\cite{Merchiers2007_PRA}, which has been widely applied for various particle clusters, including dimers~\cite{YAN_Nat.Commun._magnetically_2015,VANDEGROEP_Optica_direct_2016,SHIBANUMA_ACSPhotonics_experimental_2017,BARREDA_Nat.Commun._electromagnetic_2017}, trimers~\cite{YAN_Nat.Commun._magnetically_2015,LU2015Laser&PhotonicsReviews,BANZER_Nat.Commun._chiral_2016,YAO_ACSPhotonics_controlling_2016}, quadrumers~\cite{YAN_Nat.Commun._magnetically_2015,HOPKINS_ACSPhotonics_interplay_2015}, and other types of oligomers~\cite{Miroshnichenko2012_NL6459}. Based on the coupled-dipole theory, it is also possible to calculate directly the eigenmodes of the whole cluster~\cite{CAO_NanoLett._optical_2011,HOPKINS_ACSPhotonics_interplay_2015,VANDEGROEP_Optica_direct_2016}, making it possible to treat the whole cluster effectively as an individual scattering particle, and then the basic principles discussed above in Secs.~\textbf{2} and \textbf{3} can be directly applied.

\begin{figure*}
\centerline{\includegraphics[width=13cm]{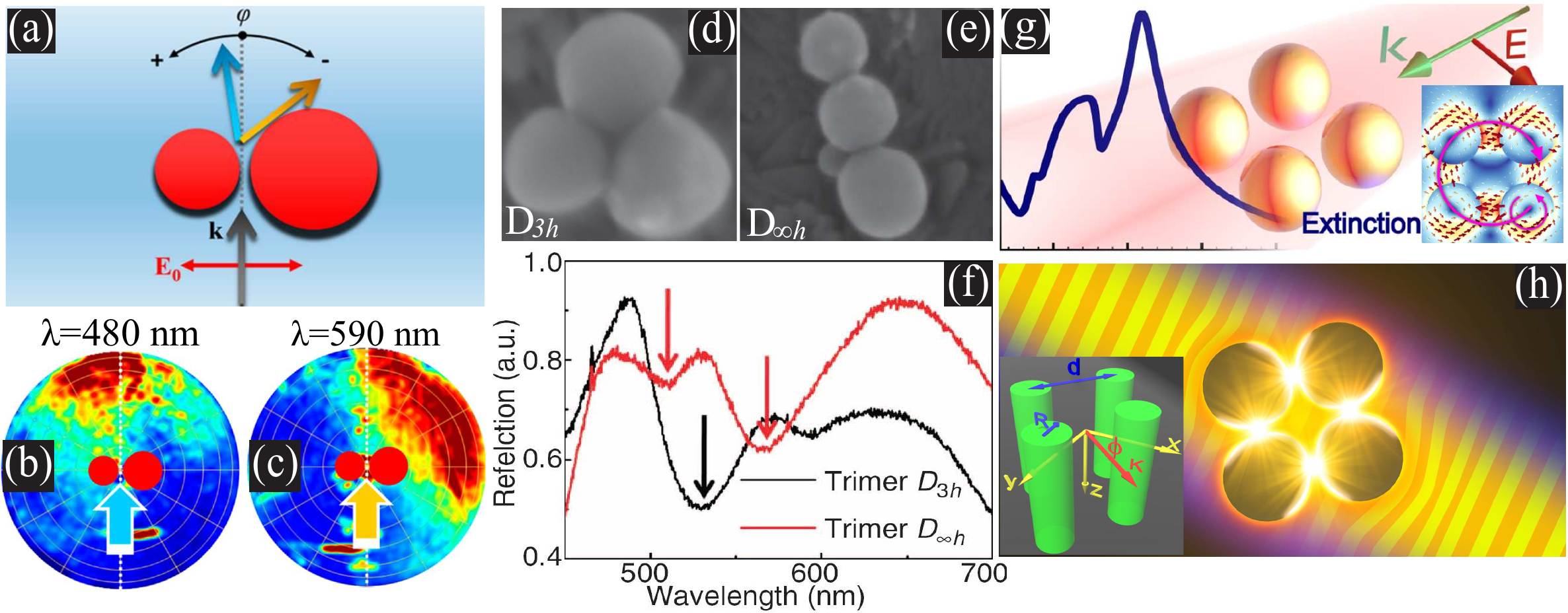}}\caption{\small Scattering manipulations for different clusters of partices, including
(a-c) asymmetric silicon dimers, (d-f) silicon trimers, and (g,h) higher-index dielectric quadrumers. The dimer can effectively direct light of different wavelengths to different directions [(b,c)], and the backscattering of the trimers (both line-shaped and triangle-shaped) can be significantly suppressed [(d-f)]. Besides angular scattering controls in (a-d), the total scattering can be significantly suppressed for the quadrumer made of high-index dielectric spheres shown in (g), or fully eliminated for that consisting of high-index dielectric cylinders shown in (h), where simultaneous free-space field enhancement has also been obtained.  Adapted from [90,92,97,101].}
\label{fig4}
\end{figure*}

Figures~\ref{fig4}(a)-\ref{fig4}(c) show the efficient routing for light of different colors to different directions based on an asymmetric silicon dimer, which is induced mainly by the interferences of the two MDs excited within each particle~\cite{SHIBANUMA_ACSPhotonics_experimental_2017}. Similar studies has also been conducted for trimers and oligomers with more consisting particles, and Figs.~\ref{fig4}(d)-\ref{fig4}(f) show that for silicon trimers (both triangle-shaped and line-shaped), effective backscattering suppression can be obtained at various wavelengths~\cite{YAN_Nat.Commun._magnetically_2015}. In a similar way, this originates from the interferences of ED and MD modes excited within all silicon particles. Besides the manipulations of angular scattering, the multipolar interference principles can also be applied to control the total scattering cross sections of particles clusters, with two examples of quadrumers shown in Figs.~\ref{fig4}(g) and \ref{fig4}(h). Figure~\ref{fig4}(g) shows the effective total scattering suppression of a quadrumer made of dielectric spheres, which comes from the destructive interference (at the Fano dip) of two eigenmodes of the whole quadrumer, both of which are mainly of magnetic nature ~\cite{HOPKINS_ACSPhotonics_interplay_2015,LIMONOV_NatPhoton_fano_2017}. A further step is made with a quadrumer consisting of high-index dielectric cylinders [see Fig.~\ref{fig4}(h)], and it is shown that the scattering can be totally eliminated, making the quadrumer effectively invisible~\cite{LIU_ArXivPrepr.ArXiv170406049_scattering_2017}. Together with invisibility, the fields are also enhanced in the accessible free space between the cylinders, which has been made possible by efficient excitations of not only ED and MD modes, but also higher-order multipoles within each cylinder~\cite{LIU_ArXivPrepr.ArXiv170406049_scattering_2017}. It is worth noticing that when higher-order multipoles within either particle are effectively excited (as is the case discussed in ~\cite{LIU_ArXivPrepr.ArXiv170406049_scattering_2017}), the coupled dipole theory can not be simply applied, and more general multiple scattering theory should be employed. Moreover, for such a system it is not convenient or direct to conduct an eigenmode analysis for the whole cluster (as has been done in ~\cite{HOPKINS_ACSPhotonics_interplay_2015}), because here you have to consider interactions not only between ED and MD modes, but also between higher-order multipoles.

Here we have confined our discussions to the clusters made of a few particles (say, up to four), but similar multi-particle multipolar interference principles can be certainly applied to clusters made of more particles and other geometrical patterns. Compared to individual particles, the interparticle interaction renders an extra dimension of freedom for manipulations of multipolar interferences, based on which more sophisticated and advanced optical functionalities related to the generalized Kerker effects could be obtained.

\section{Generalized Kerker effects for metalattices}

Finally, we discuss the generalized Kerker effects for extended periodic structures composed of individual particles or periodically arranged particle clusters. Here we employ the term ``metalattices", instead of commonly used term ``metasurfaces" or less frequent term``metagratings", as we cover both the metasurface regime of subwavelength periodicity where there exists no higher-order diffraction (transmission and reflection are viewed as the zeroth-order diffractions), and the metagrating regime of larger periodicity when efficient coupling to higher-order diffractions is observed. Traditionally, the prefix ``meta" is used to differentiate the periodic structures we discuss here from the conventional periodic structures (such as plasmonic structures) that can be described mainly by the ED modes. Within each unit cell of the metalattice, not only ED modes but also MD modes or higher-order multipoles can be efficiently excited, which is similar to the cases considered in Secs.~\textbf{2}-\textbf{4}.

Generally speaking, a \textit{metalattice} is a special type of extended cluster of scattering particles. Compared to the finite particle cluster that has potentially infinite out-coupling channels (scattering at different angles, which is constrained by the optical theorem though~\cite{hulst_light_1957,jackson1962classical,Bohren1983_book}), for metalattices there appear only finite diffraction orders (out-coupling channels).  When loss is involved, the system absorption can be viewed as an extra out-coupling channel for both cases~\cite{HAUS_waves_1984}. Similarly, the multiple scattering theory can also be applied to metalattices~\cite{QUINTEN_optical_2010}, through which the multipolar excitation efficiencies within each unit cell can be directly calculated, with the lattice coupling effects simultaneously taken into account~\cite{DEABAJO_Rev.Mod.Phys._colloquium_2007,BYELOBROV_IEEEAntennasPropag.Mag._periodicity_2015}.  When each unit cell (or each consisting particle within a unit cell made of particle clusters) can be effectively simplified as dipoles (including both ED and MD modes), the simpler coupled dipole theory can be applied directly~\cite{Merchiers2007_PRA,Liu2012_PRB}. The analysis of metalattices can be simplified significantly by applying the Floquet theory~\cite{Joannopoulos2008_book}, where one may calculate the multipolar excitations within single rather than all unit cells. Then the problem is reduced to a scattering problem for one unit cell, that is even simpler for metalattices than for individual particles or particle clusters, as for metalattices we should consider only the scattering along a finite number of diffraction directions besides the system absorption.

Below,  we review the generalized Kerker effects employed to manipulate various out-coupling channels of metalattices,  including transmission, reflection, absorption, as well as higher-order diffraction. We focus on the perfect transmission, perfect reflection, perfect absorption, and large-angle beam routing based on sophisticated diffraction manipulations, all of which have been enabled through the interferences of multipoles excited within metalattices.

\subsection{Perfect transmission}

To achieve perfect transmission, one should exclude all other possible out-coupling scattering channels. This is easier to achieve with metasurfaces without higher-order diffractions, as only reflection elimination is required if the structure is lossless.  From the fundamental point of view, the sufficient requirement to achieve this condition is to fully suppress the backward scattering from each unit cell with normally incident waves, which has already been discussed extensively above in Secs.~\textbf{2}-\textbf{4} above (see also Fig.~\ref{fig1}).

\begin{figure*}
\centerline{\includegraphics[width=13cm]{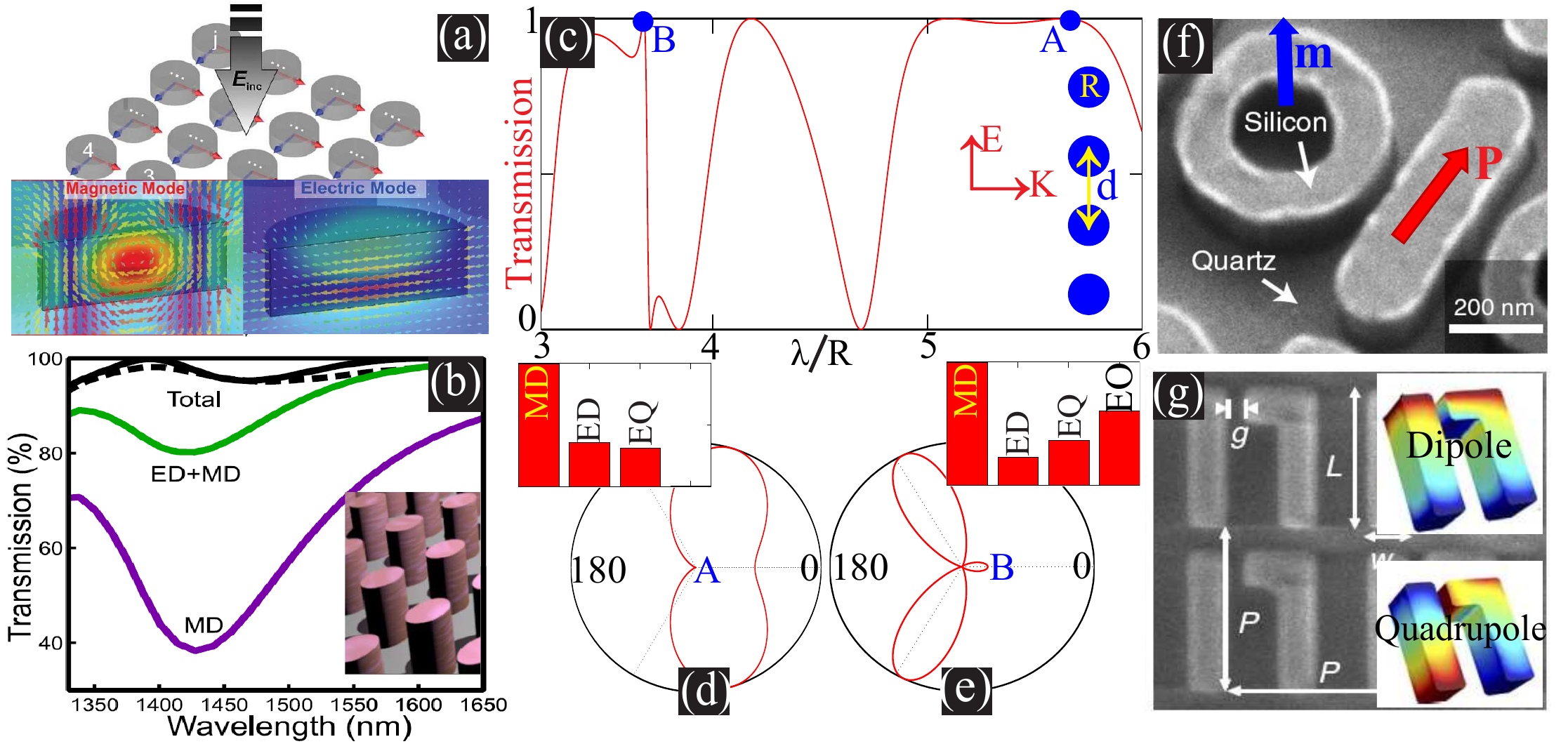}}\caption{\small Perfect transmission for lattices of Si disks where the reflection is eliminated by interference of: ED and MD modes in (a); ED, MD and other unspecified higher-order multipoles in (b). Similarly, a 1D lattice of high-index dielectric cylinders [see inset in (c)] can also be made fully transparent at the points \textbf{A} and \textbf{B} indicated in (c), which originate respectively from the interferences of: ED, MD, and EQ modes shown in (d); ED, MD, EQ, and EO modes shown in (e). In (c) and (d), both the angular scattering patterns of each unit cell and relative total scattered power of all excited multipoles are shown. Perfect transmission can also be obtained with metalattice consisting of dimers as the unit cells, as is shown in (f) and (g). Reflection is fully suppressed by interference of ED and MD resonances in (f), and of ED and EQ resonances in (g). Adapted from [106,113,118-120].}
\label{fig5}
\end{figure*}

Figure~\ref{fig5} summarizes the demonstrations of the perfect transmission with metalattices made of individual particles [Figs.~\ref{fig5}(a)-\ref{fig5}(e)] or particle dimers [Figs.~\ref{fig5}(f) and \ref{fig5}(g)]. For the case of individual particles: (i) Figure~\ref{fig5}(a) shows the most widely employed scenario of Huygens' source with overlapped ED and MD resonances~\cite{decker_high-efficiency_2015} in a specially-designed metalattice termed as ``Huygens' surface"~\cite{KUZNETSOV_Science_optically_2016,LIU_ArXivPrepr.ArXiv160901099_multipolar_2016,STAUDE_NatPhoton_metamaterialinspired_2017,CHEN_Rep.Prog.Phys._review_2016,KIVSHAR_Opt.PhotonicsNews_metaoptics_2017,
KRUK2017ACSPhotonics,YANG2017PhysicsReports,DING_Rep.Prog.Phys._gradient_2017,KIM_Phys.Rev.X_optical_2014,WANG_Appl.Phys.Lett._generation_2014,LIU_NanoLett._huygens_2017-1,
BABICHEVA_Laser&PhotonicsReviews_resonant,VASILANTONAKIS_ArXiv171101430Phys._designing_2017,CHENKE_2017_AdvancedMaterials_Reconfigurable}. (ii) As discussed in Sec.~\textbf{1}, the zero backward scattering for individual particles can also be obtained with more interfering components involving higher-order multipoles.  This is  demonstrated in Fig.~\ref{fig5}(b) with a metalattice of Si disks, though it is not specified exactly what the higher-order multipoles involved are~\cite{KRUK_APLPhotonics_invited_2016}. (iii) Several other studies have identified almost all the multipoles that have been involved~\cite{Spinelli2012_NC,RODRIGUEZ_Phys.Rev.Lett._breaking_2014,CAMPIONE_Opt.ExpressOE_tailoring_2015,YANG_Phys.Rev.B_multimode_2017-1}, while unfortunately their analysis is based on isolation of the unit cells, without taking the lattice coupling effects into consideration. (iv) A more comprehensive and thorough study is shown in Figs.~\ref{fig5}(c)-\ref{fig5}(e), where the perfect transmission is achieved with a metalattice consisting of an array of high-index cylinders [see the inset in Fig.~\ref{fig5}(c), with the incident wave polarized along the array]~\cite{liu_beam_2017}. It is shown convincingly that, the perfect transmission at the point A comes from the interferences of MD, ED, and EQ resonances [see Fig.~\ref{fig5}(d), where the angular scattering pattern of the unit cell and the scattered power of all involving multipoles are shown], while at the point B an extra multipole (electric octupole, EO) will join, with the same core feature of the zero backward scattering of the unit cell preserved [see Fig.~\ref{fig5}(e)].

According to the discussions in Sec.~\textbf{4}, multipolar interferences can also eliminate the backscattering of particle dimers. As a result, the perfect transmission can also be obtained with metalattices made of dimer-type unit cells. Two such examples are shown in Figs.~\ref{fig5}(f) and \ref{fig5}(g), where the perfect transmission originates respectively from interferences mainly between ED and MD in ~\cite{YANG_NatCommun_alldielectric_2014}, ED and EQ in ~\cite{WU_NatCommun_spectrally_2014}. Similarly, other combinations of multipoles can also render the whole metalattice fully transparent~\cite{ZHAO_Opt.ExpressOE_dipole_2014}. The involving multipoles are spatially separated as they are supported by different particles within each unit cell in Fig.~\ref{fig5}(f). These studies can certainly be extended to obliquely incident waves~\cite{liu_beam_2017,PANIAGUA-DOMINGUEZ_NatCommun_generalized_2016-1}, and to metalattices made of more complicated unit cells, consisting of more particles of other distributions, and involving more multipoles of higher orders. Here we discuss only the regime of metasurfaces, and for metagratings it is more challenging to obtain perfect transmission, as not only reflection, but also all other higher-order diffractions have to be fully suppressed. Almost perfect (more than 95\%) transmission has been demonstrated with metalattices (metagratings) consisting of high-index particles, where multipolar interferences are engineered to significantly suppress all other out-coupling channels except transmission~\cite{liu_beam_2017}.

\subsection{Perfect reflection}

\begin{figure*}
\centerline{\includegraphics[width=13cm]{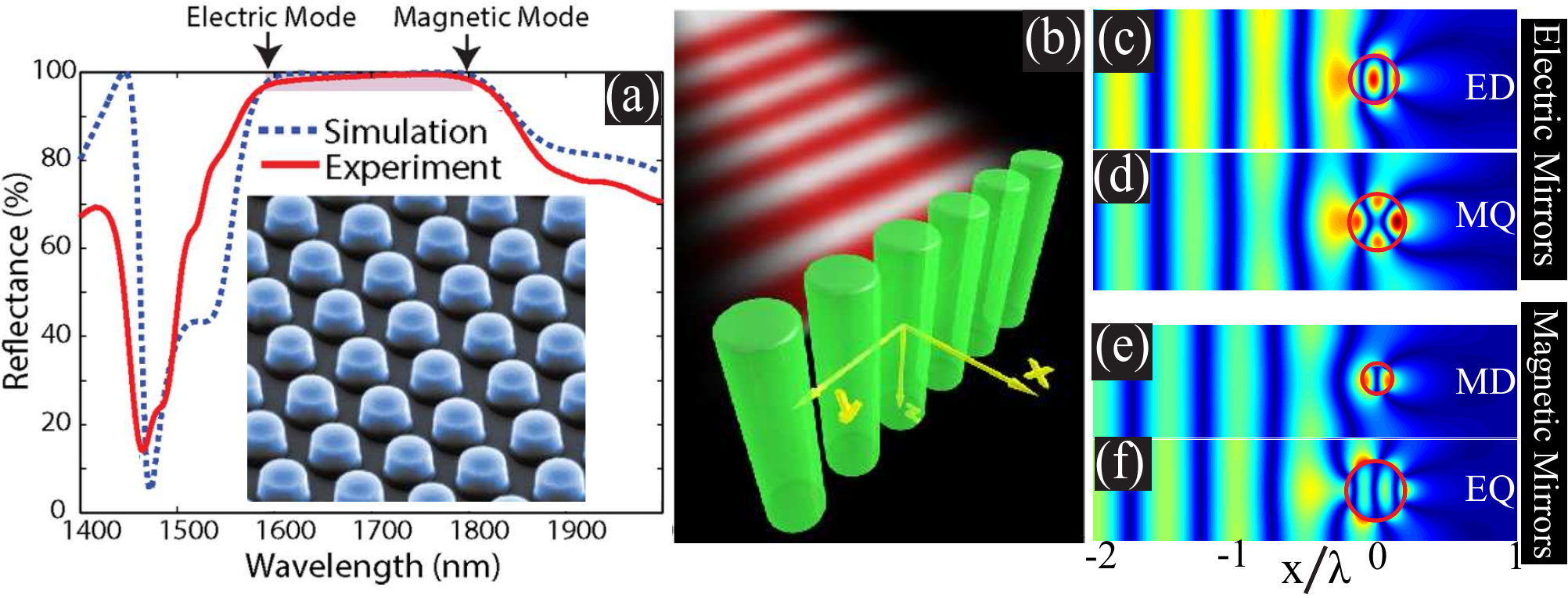}}\caption{\small (a) A metalattice of Si nanodisks (inset) reflects the incident wave at ED or MD resonances. (b) Perfect reflection by a lattice of high-index dielectric cylinders. Electric mirrors are induced by  ED or MQ resonances [(c-d)], and magnetic mirrors induced by MD or EQ resonances [(e-f)] within each lattice cylinder. Adapted from [24,126].}
\label{fig6}
\end{figure*}

Similar to the case of perfect transmission, it is easier to realize the perfect reflection in the metasurface regime, where only the transmission elimination is required. In a sharp contrast to full reflection suppression that relies on the zero backward scattering of each unit cell, transmission elimination is induced by significant forward scattering to interfere destructively with the incident wave. In many cases, the perfect reflection is associated with resonant multipolar excitations within each unit cell of the metalattices, and the most widely studied case is the resonant excitation of ED and/or MD modes~\cite{GHENUCHE_Phys.Rev.Lett._optical_2012,DU_Phys.Rev.Lett._nearly_2013,SLOVICK2013Phys.Rev.B,LIU_OpticaOPTICA_optical_2014,WANG_Appl.Phys.Lett._generation_2014,
MOITRA_ACSPhotonics_largescale_2015,shalaev_high-efficiency_2015,LI_2017_ACSNano_Dielectric}.
One such example is shown in Fig.~\ref{fig6}(a), where a metalattice of Si nanodisks can fully reflect the incident wave with resonant ED or MD modes within each unit cell. Similar effect can be achieved with the higher-order multipoles~\cite{LIU_Phys.Rev.Lett._generalized_2017,liu_beam_2017}, where quadrupole and/or octupole resonances (or their combinations with ED and MD modes) can also result in full wave reflection.

Perfect reflection can be further categorized based on the phase (relative to the incident wave) of the reflected waves, which are highly related to the recent studies of electric and magnetic mirrors~\cite{LIU_OpticaOPTICA_optical_2014,MOITRA_ACSPhotonics_largescale_2015,LIU_Phys.Rev.Lett._generalized_2017}. For an electric mirror, there is a half-wave loss (in terms of electric field) of the reflected wave, whereas for a magnetic mirror there is no such phase jump, and other cases can be placed between them. Based on the phase symmetries of multipoles of different natures and orders (as shown in Fig.~\ref{fig1}), a recent study provides a comprehensive dictionary that establishes a full correspondence between multipoles and mirrors~\cite{LIU_Phys.Rev.Lett._generalized_2017}. It is shown that the type of mirrors obtained is solely decided by the phase pariy rather than the order or nature of the multipoles excited. One simple example is shown in Fig.~\ref{fig6}(b), where the incident wave can be fully reflected by a lattice of high-index dielectric cylinders. Resonant excitation of ED or MQ modes that show even parity leads to electric mirrors [see Figs.~\ref{fig6}(c) and \ref{fig6}(d)], whereas MD and EQ resonances exhibit odd parity which results in magnetic mirrors [Figs.~\ref{fig6}(e) and \ref{fig6}(f)]. This is rather counterintuitive since magnetic (electric) mirrors do not necessarily require the existence of magnetic (electric) resonances.

\begin{figure*}
\centerline{\includegraphics[width=13cm]{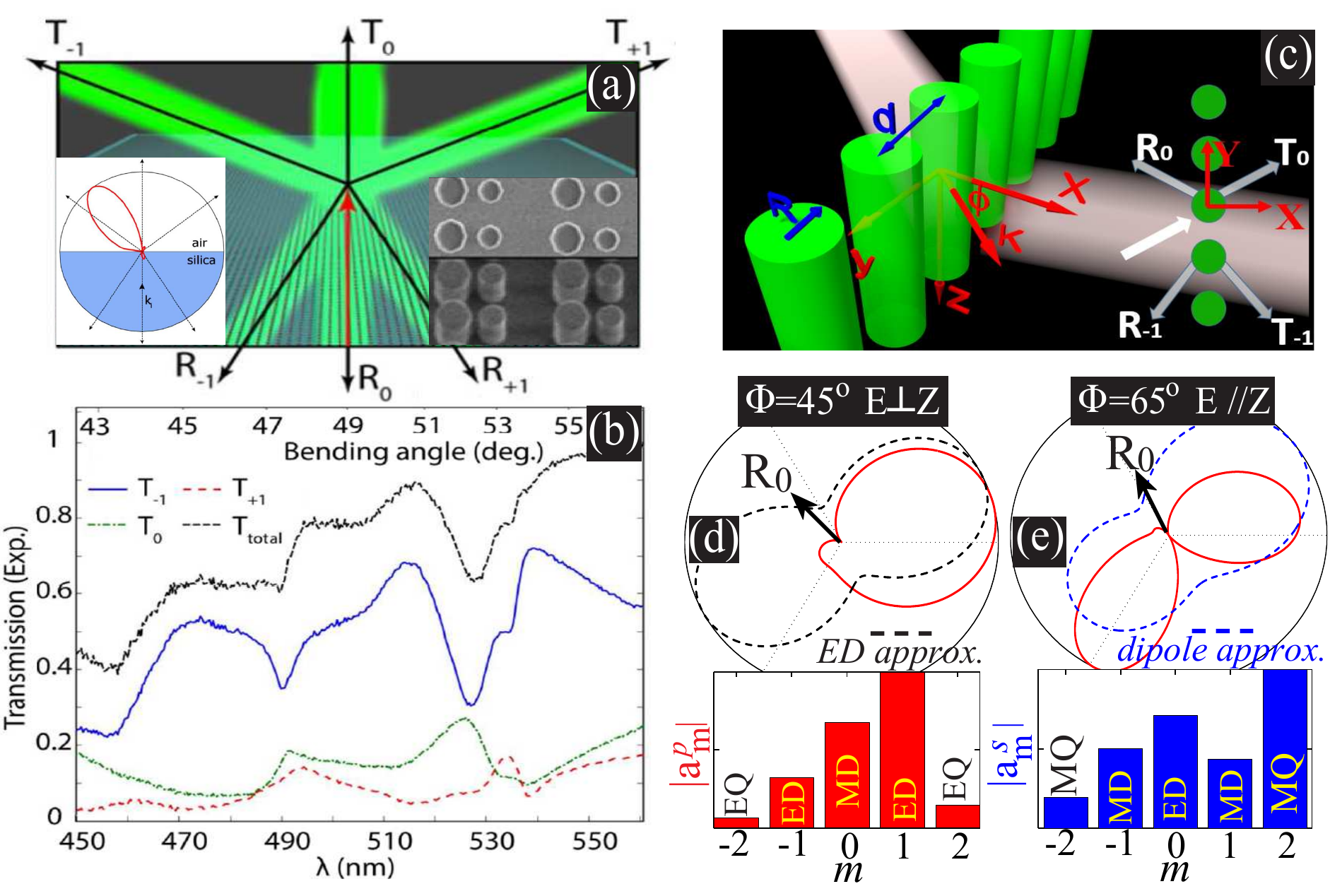}}\caption{(a) A metalattice made of asymetric $\rm{TiO_2}$ dimers (shown in the right inset) can direct most of the transmitted light to the ($-1$) order diffraction, thus realising effectively highly efficient large-angle beam bending at various wavelengths, as shown in (b). This has been enabled by the high directional scattering of each unit cell [left inset in (a)]. (c) With a much simpler 1D metalattice made of high-index dielectric cylinders, almost the same functionality can be obtained for both polarizations of incident waves. This is induced by the specific scattering patterns of the lattice cylinders [upper row in (d, e)], which originate from interferences of multipoles of different orders [lower row in (d, e); it agrees with the upper row that the dipolar approximation (dashed curves) is not sufficient here and higher-order multipoles should be taken into consideration].  Adapted from [118,134].}
\label{fig7}
\end{figure*}

\subsection{Higher-order diffraction control}

Metalattices operating in the metagrating regime have the lattice spacing larger (or comparable to) the incident wavelength, for which higher-order diffraction effects naturally appear. Conventional studies of gratings emphasize greatly their collective responses such as grating diffractions, where the effect of lattice couplings on scattering pattern of each unit cell is largely neglected. As a result, it is widely assumed that lower-order diffraction effects should be stronger than those higher-order ones. Nevertheless, the grating diffraction itself just means that the scattered light interferes constructively along the corresponding diffraction direction. If the scattering of the unit cell along a diffraction angle is fully eliminated, there will be no energy routed into this diffraction order. Basically, we need to consider not only the collective grating diffractions, but also the scattering pattern of each unit cell (with the lattice coupling effects considered) to predict accurately by what proportion the incident wave will be channelled to  different diffraction directions. Consequently, as generalized Kerker effects can be employed to shape the angular scattering
of each unit cell, the principles of multipolar interference can certainly provide extra opportunities for diffraction management in meta-optics.

Higher-order diffraction management in metalattices has been demonstrated in the dipolar regime (where only  interference between ED and MD modes is employed to shape the unit-cell scattering~\cite{DU_Phys.Rev.Lett._optical_2011,WU_NanoLett._experimental_2015,RADI_Phys.Rev.Lett._metagratings_2017,CHALABI_Phys.Rev.B_efficient_2017}), and also in more general multipolar regimes where higher-order multipoles are taken into account~\cite{PANIAGUA-DOMINGUEZ_ArXiv170500895Phys._metalens_2017,KHAIDAROV_NanoLett._asymmetric_2017,liu_beam_2017}. Two such examples with higher-order interfering multipoles are shown in Fig.~\ref{fig7}. Figures~\ref{fig7}(a) and \ref{fig7}(b) show the results for a metalattice made of asymmetric $\rm{TiO_2}$ dimers [right inset in Fig.~\ref{fig7}(a)]. For various wavelengths, a great proportion of the incident light is channelled into the ($-1$) order diffraction of transmission, and all other possible diffraction channels are suppressed significantly, obtaining effectively the functionality of large-angle beam bending~\cite{KHAIDAROV_NanoLett._asymmetric_2017}. This functionality is made possible by highly directional scattering of each unit cell [left inset in Fig.~\ref{fig7}(a)], though it was not specified exactly what the involved higher-order multipoles are and how efficiently they have been excited. It is worth mentioning that such functionality has been widely discussed in various gradient metasurface geometries~\cite{DING_Rep.Prog.Phys._gradient_2017}, where however it is rather challenging to achieve sufficiently high efficiency.

Similar functionalities can also be achieved with metalattices of much simpler designs, and one example is shown Fig.~\ref{fig7}(c). Basically, a lattice made of high-index dielectric cylinders can significantly bend the incident waves [of both in-plane and out-of-plane polarizations; see inset in Fig.~\ref{fig7}(c)] by a large angle with high efficiencies~\cite{liu_beam_2017}. Figures~\ref{fig7}(d) and \ref{fig7}(e) show the corresponding angular scattering patterns for the lattice cylinder (see the upper row), revealing clearly also what multipoles are excited [see the lower row in Figs.~\ref{fig7}(d) and \ref{fig7}(e)] to interference with one another to produce such scattering patterns. For both polarizations, the lattice cylinders do not scatter at the reflection direction, but scatter significantly in the forward direction to suppress the transmission. Undoubtedly, similar studies can be conducted for metalattices made of more complicated unit cells, and of larger periodicity when there are extra higher-order diffractions.

\subsection{Perfect absorption}

The discussion of metalattices presented above is limited to lossless structures. When losses
are involved, we have to consider, not only the diffractions (including reflection and transmission),
but also the absorption of the whole lattice. Generally speaking, losses would change the magnitudes and phases of all multipoles excited within each unit cell, especially near the resonances. Under this circumstance, both the phase parities and the absorption-induced phase differences between different multipoles have to be considered simultaneously. They would affect the angular scattering patterns and consequently the energy distributions among all possible diffraction orders of the whole lattice. As a result, we anticipate that losses would provide extra freedom for the multipolar tuning (in
terms of both phases and magnitudes) and thus allow for more flexible beam manipulations.

\begin{figure*}
\centerline{\includegraphics[width=12cm]{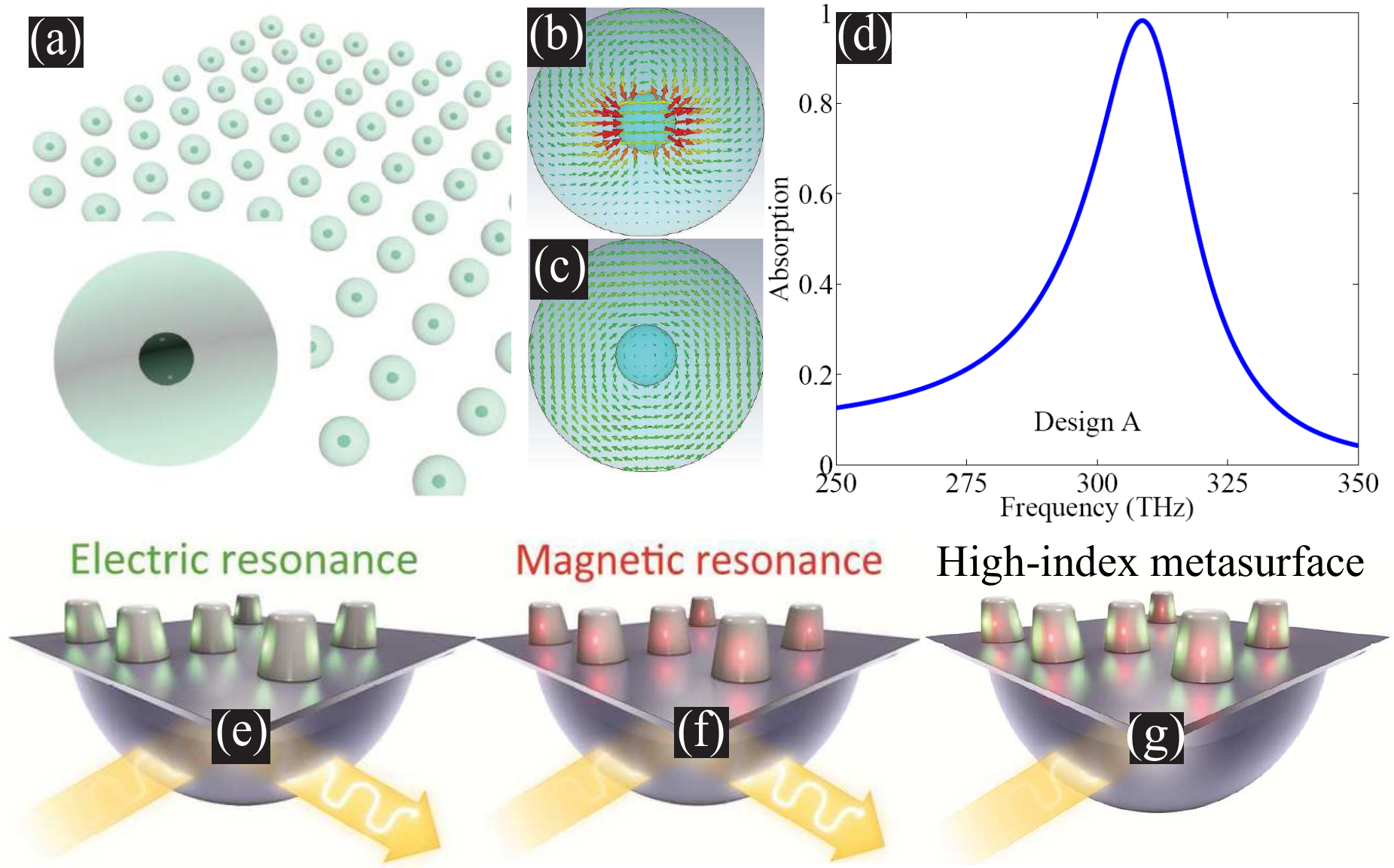}}\caption{\small Perfect absorption associated with generalized Kerker effects. A 2D metalattice made of core-shell spheres shown in (a) can fully absorb incident
waves shone from one side, as shown in (d). The core-shell particle simultaneously support
both ED and MD resonances, with the corresponding near field distributions shown in (b,c). (e-g) Total internal reflection with dielectric particles. (g) When the particle supports both ED and MD resonances, the reflection can be fully eliminated, achieving  the perfect absorption of the incident wave. However, if the particle support either ED or MD resonances only, the incident wave will be absorbed only partially, as shown in (e,f). Adapted from [139,140].}
\label{fig8}
\end{figure*}

An extreme case of the loss-induced beam control is that all the diffractions of the metalattices can be effectively eliminated, with all the incident wave perfectly absorbed~\cite{WATTS_Adv.Mater._metamaterial_2012,RADI_Phys.Rev.Appl._thin_2015,BARANOV2017Nat.Rev.Mater.}. Current studies to obtain the perfect absorption with metalattices rely dominantly on the interferences of ED and MD modes, which can eliminate reflection~\cite{LAROCHE_ArXivPrepr.Physics0606119_controlling_2006,WATTS_Adv.Mater._metamaterial_2012,RADI_Phys.Rev.Appl._thin_2015,RADI_ACSPhotonics_full_2015,ODEBOLANK_NanoLett._largescale_2017,
ALAEE_ArXiv171108203Phys._theory_2017,YANG_2018_ACSPhotonics}.
Two such scattering configurations are presented in Fig.~\ref{fig8}, where the perfect absorption can be achieved~\cite{RADI_ACSPhotonics_full_2015,ODEBOLANK_NanoLett._largescale_2017}. Figure~\ref{fig8}(d) shows the perfect absorption by a metalattice made of metal-dielectric core-shell particles [shown in Fig.~\ref{fig8}(a)]~\cite{RADI_ACSPhotonics_full_2015}. The core-shell particle supports a pair of overlapped ED and MD resonances [shown in Figs.~\ref{fig8}(b) and \ref{fig8}(c)]~\cite{RADI_ACSPhotonics_full_2015,Liu2012_ACSNANO}, which can eliminate simultaneously both reflection and transmission (for each particle, there is no backward scattering, and simultaneously there strong forward superscattering to interfere destructively with the incident wave), with the incident wave fully absorbed. We emphasize here that the incident wave is sent from one side, which is contrastingly different from the coherent absorption achieved with double-sided incident waves~\cite{BARANOV2017Nat.Rev.Mater.}, for which sole ED mode excitation would be sufficient. Figures~\ref{fig8}(e)-\ref{fig8}(g) show other configurations based on total internal reflections~\cite{ODEBOLANK_NanoLett._largescale_2017}. When the particles on the substrate support overlapped ED and MD resonances, the scattered wave can interfere destructively with the reflected wave, enabling a perfect absorption of the incident wave [Fig.~\ref{fig8}(g)]. In contrast, the perfect absorption is not possible if the particles support sole ED or MD resonances, where the reflected wave cannot be fully eliminated, ending up with a partial absorption [Figs.~\ref{fig8}(e) and \ref{fig8}(f)]. Compared to the functionality of high-order diffraction management discussed above in Sec.~5.3, for perfect absorption or significant absorption enhancement, the requirement for periodicity and other distribution parameters is not as strict as long as there exist no significant inter-particle interactions or strong collective responses. This is the case for the results shown in Figs.~\ref{fig8}(e)-\ref{fig8}(g), and also for several other studies~\cite{MOREAU_Nature_controlledreflectance_2012,AKSELROD_Adv.Mater._largearea_2015}. Here, we only mention the interference of ED and MD resonances for prefect absorption, but perfect absorption can be achieved through interferences of higher-order multipoles.

\vspace{12pt}

We emphasize here that many optical properties of metalattices have already been discussed in literature, including  perfect transmission~\cite{SUN_Appl.Phys.Lett._broadband_2008,CHEN_Phys.Rev.Lett._antireflection_2010,CHANG-HASNAIN_Adv.Opt.Photon.AOP_highcontrast_2012,COLLIN_Rep.Prog.Phys._nanostructure_2014,ARBABI_Nat.Nanotechnol._dielectric_2015,ZHAN_ACSPhotonics_lowcontrast_2016,lalanne_metalenses_2017,KHORASANINEJAD_Science_metalenses_2017}, perfect reflection~\cite{MATEUS_IEEEPhotonicsTechnol.Lett._ultrabroadband_2004,BRUCKNER_Phys.Rev.Lett._realization_2010,FATTAL_NatPhoton_flat_2010,CHANG-HASNAIN_Adv.Opt.Photon.AOP_highcontrast_2012,YANG_NanoLett._dielectric_2014,
ESFANDYARPOUR_Nat.Nanotechnol._metamaterial_2014,COLLIN_Rep.Prog.Phys._nanostructure_2014,lalanne_metalenses_2017,ARBABI_Nat.Photonics_planar_2017}, high-order diffraction management~\cite{ZHU_Optica_flexible_2015,LIN_Sci.Rep._optical_2017}, and perfect absorption~\cite{AYDIN_Nat.Commun._broadband_2011,WATTS_Adv.Mater._metamaterial_2012,COLLIN_Rep.Prog.Phys._nanostructure_2014,brongersma2014light,BARANOV2017Nat.Rev.Mater.}. However, for most cases no explicit connections between those functionalities and the generalized Kerker effects have been established. The concept of multipolar interference behind the generalized Kerker effects provides an intuitive and powerful theoretical framework that can be applied not only to explain exotic optical phenomena, but also to provide a guidance to novel designs of optical metadevices for advanced photonic applications.

\section{Perspectives and outlook}

We have provided a coherent overview of the generalized Kerker effects and their manifestations in nanophotonics and meta-optics, largely driven by resonant excitation and interference of different electromagnetic multipoles in subwavelength photonic structures. Based on the phase symmetry analysis for multipoles of different orders and origins, we have presented a general analysis for multipolar interference, based on which we have discussed various optical properties of individual particles, particle clusters, and metalattices. For finite systems of individual and clustered particles, we have focused on the control of both angular and total scatterings, whereas for metalattices we have discussed the effects of perfect reflection, transmission and absorption, as well as high-order diffraction management. We have demonstrated that all those optical phenomena are linked directly to the multipolar interference mechanism of the generalized Kerker effects.

Further progress of this field is expected through a number of potential extensions and generalizations, and below we list just a few promising ideas. (i) Almost all previous studies have only explored interference of low-order multipoles (up to quadrupoles), whereas higher-order multipoles can certainly bring extra opportunities.  (ii) Many current investigations of angular scattering focused only on the forward and backward directions, while a full-angular analysis (in terms of both amplitude and phase) for multipolar interferences is vitally important. (iii) A large amount of work in this field relies on plane-wave or dipole-emitter excitations, and there will be much more flexibilities for the wave manipulation with spatially engineered electromagnetic sources with spin and/or angular momentum (the so-called structured light), or other types of excitation configurations, such as excitations by electron beams. (iv) Very few works address the near-field Kerker effects and related multipolar interferences (see Fig.~\ref{fig3}), for which the most significant step is to identify comprehensively all possible field components (both electric and magnetic ones) and their phase symmetries for the multipoles (especially higher-order ones) in the near field. (v) The field of active resonant dielectric nanophotonics is rapidly developing, so it is crucial to expand these studies to the case of active particles, where such effects as gain and chirality can play an important role. (vi) Last but not least, the study of periodic lattices can be extended to other aperiodic geometries, including quasi-periodic or random lattices.

With those further important extensions, we anticipate that the  generalized Kerker effects discussed above can emerge in other fields
and also affect other advanced and intriguing photonic effects, including the quantum~\cite{TAME_NatPhys_quantum_2013}, low-dimensional~\cite{xia_twodimensional_2014,KHANIKAEV2017Nat.Photonics} and topological~\cite{LU_NatPhoton_topological_2014,KHANIKAEV2017Nat.Photonics} systems, as well as make an impact on the studies of
geometric phases~\cite{LI_Nat.Rev.Mater._nonlinear_2017,MAGUID2017Science}, bandgap structures~\cite{RYBIN_Nat.Commun._phase_2015}, Anderson localization~\cite{segev2013anderson}, and non-Hermitian photonics~\cite{FENG2017Nat.Photonics}, also proliferating the applications
ranging from the scattering control and wave guidance~\cite{BARREDA_SciRep_light_2017} to more advanced phenomena including optical force control~\cite{GAO_LightSci.Appl._optical_2017}, thermal emission engineering~\cite{ZHAI_Science_scalablemanufactured_2017-1}, quantum interference and entanglement management~\cite{JHA2015Phys.Rev.Lett.,JHA2017ACSPhotonics}, in both linear and nonlinear regimes~\cite{LI_Nat.Rev.Mater._nonlinear_2017,SMIRNOVA_Optica_multipolar_2016}. It is also significant to establish connections between the scattering of electromagnetic waves and scattering of other types of waves from the perspective of partial wave interferences~\cite{CUMMER2008Phys.Rev.Lett.,ZHANG2008Phys.Rev.Lett.,FARHAT2008Phys.Rev.Lett.,LIAO2012Phys.Rev.Lett.,BRULE2014Phys.Rev.Lett.,CLOS2016Phys.Rev.Lett.}, so that the principles of generalized Kerker effects could be explored in other fields, inspiring interdisciplinary studies
and practical applications.

\section*{Funding}
National Natural Science Foundation of China (Grant 11404403); Outstanding Young Researcher Scheme of National University of Defense Technology of China;  Australian Research Council.

\section*{Acknowledgments}
We thank our numerous co-authors and colleagues for their input, suggestions, and productive collaborations.


\begin{thebibliography}{100}
\newcommand{\enquote}[1]{``#1''}

\bibitem{Kerker1983_JOSA}
M.~Kerker, D.~S. Wang, and C.~L. Giles, \enquote{Electromagnetic scattering by
  magnetic spheres,} J. Opt. Soc. Am. \textbf{73}, 765 (1983).

\bibitem{hulst_light_1957}
H.~C. Hulst, \emph{Light Scattering by Small Particles} (Courier Corporation,
  1957).

\bibitem{jackson1962classical}
J.~D. Jackson, \emph{Classical Electrodynamics} (Wiley New York, 1962).

\bibitem{Bohren1983_book}
C.~F. Bohren and D.~R. Huffman, \emph{Absorption and Scattering of Light by
  Small Particles} (Wiley, 1983).

\bibitem{Jin2010_IEEE}
P.~Jin and R.~W. Ziolkowski, \enquote{Metamaterial-inspired, electrically small
  {{Huygens'}} sources,} Antennas and Wireless Propagation Letters, IEEE
  \textbf{9}, 501 (2010).

\bibitem{Pendry1999_ITMT}
J.~B. Pendry, A.~J. Holden, D.~J. Robbins, and W.~J. Stewart,
  \enquote{Magnetism from conductors and enhanced nonlinear phenomena,} IEEE.
  T. Microw. Theory \textbf{47}, 2075 (1999).

\bibitem{Zhao2009_materialtoday}
Q.~Zhao, J.~Zhou, F.~Zhang, and D.~Lippens, \enquote{Mie resonance-based
  dielectric metamaterials,} Mat. Today \textbf{12}, 60 (2009).

\bibitem{Kuznetsov2012_SciRep}
A.~I. Kuznetsov, A.~E. Miroshnichenko, Y.~H. Fu, J.~B. Zhang, and B.~S.
  Lukyanchuk, \enquote{Magnetic light,} Sci. Rep. \textbf{2}, 492 (2012).

\bibitem{Evlyukhin2012_NL}
A.~B. Evlyukhin, S.~M. Novikov, U.~Zywietz, R.~L. Eriksen, C.~Reinhardt, S.~I.
  Bozhevolnyi, and B.~N. Chichkov, \enquote{Demonstration of magnetic dipole
  resonances of dielectric nanospheres in the visible region,} Nano Lett.
  \textbf{12}, 3749 (2012).

\bibitem{Liu2014_CPB}
W.~Liu, A.~E. Miroshnichenko, and Y.~S. Kivshar, \enquote{Control of light
  scattering by nanoparticles with optically-induced magnetic responses,} Chin.
  Phys. B \textbf{23}, 047806 (2014).

\bibitem{CHEN_Rep.Prog.Phys._review_2016}
H.-T. Chen, A.~J. Taylor, and N.~Yu, \enquote{A review of metasurfaces: physics
  and applications,} Rep. Prog. Phys. \textbf{79}, 076401 (2016).

\bibitem{KUZNETSOV_Science_optically_2016}
A.~I. Kuznetsov, A.~E. Miroshnichenko, M.~L. Brongersma, Y.~S. Kivshar, and
  B.~Luk'yanchuk, \enquote{Optically resonant dielectric nanostructures,}
  Science \textbf{354}, aag2472 (2016).

\bibitem{LIU_ArXivPrepr.ArXiv160901099_multipolar_2016}
W.~Liu and Y.~S. Kivshar, \enquote{Multipolar interference effects in
  nanophotonics,} Phil. Trans. R. Soc. A \textbf{375}, 20160317 (2017).

\bibitem{STAUDE_NatPhoton_metamaterialinspired_2017}
I.~Staude and J.~Schilling, \enquote{Metamaterial-inspired silicon
  nanophotonics,} Nat. Photonics \textbf{11}, 274 (2017).

\bibitem{KIVSHAR_Opt.PhotonicsNews_metaoptics_2017}
Y.~Kivshar and A.~Miroshnichenko, \enquote{Meta-{Optics} with {{Mie}}
  resonances,} Opt. Photonics News \textbf{28}, 24 (2017).

\bibitem{KRUK2017ACSPhotonics}
S.~Kruk and Y.~Kivshar, \enquote{Functional meta-optics and nanophotonics
  governed by {{Mie}} resonances,} ACS Photonics \textbf{4}, 2638 (2017).

\bibitem{YANG2017PhysicsReports}
Z.-J. Yang, R.~Jiang, X.~Zhuo, Y.-M. Xie, J.~Wang, and H.-Q. Lin,
  \enquote{Dielectric nanoresonators for light manipulation,} Phys. Rep.
  \textbf{701}, 1 (2017).

\bibitem{DING_Rep.Prog.Phys._gradient_2017}
F.~Ding, A.~Pors, and S.~I. Bozhevolnyi, \enquote{Gradient metasurfaces: a
  review of fundamentals and applications,} Rep. Prog. Phys. \textbf{81},
  026401 (2018).

\bibitem{Radescu2002_PRE}
E.~E. Radescu and G.~Vaman, \enquote{Exact calculation of the angular momentum
  loss, recoil force, and radiation intensity for an arbitrary source in terms
  of electric, magnetic, and toroid multipoles,} Phys. Rev. E \textbf{65},
  046609 (2002).

\bibitem{liu_efficient_2015}
W.~Liu, J.~Shi, B.~Lei, H.~Hu, and A.~E. Miroshnichenko, \enquote{Efficient
  excitation and tuning of toroidal dipoles within individual homogenous
  nanoparticles,} Opt. Express \textbf{23}, 24738 (2015).

\bibitem{alaee_electromagnetic_2018}
R.~Alaee, C.~Rockstuhl, and I.~Fernandez-Corbaton, \enquote{An electromagnetic
  multipole expansion beyond the long-wavelength approximation,} Opt. Commun.
  \textbf{407}, 17 (2018).

\bibitem{liu_ultradirectional_2014}
W.~Liu, J.~Zhang, B.~Lei, H.~Ma, W.~Xie, and H.~Hu, \enquote{Ultra-directional
  forward scattering by individual core-shell nanoparticles,} Opt. Express
  \textbf{22}, 16178 (2014).

\bibitem{hulst_quantifying_2012}
N.~F.~v. Hulst, R.~Zia, S.~Karaveli, and T.~H. Taminiau, \enquote{Quantifying
  the magnetic nature of light emission,} Nat. Commun. \textbf{3}, 979 (2012).

\bibitem{LIU_Phys.Rev.Lett._generalized_2017}
W.~Liu, \enquote{Generalized magnetic mirrors,} Phys. Rev. Lett. \textbf{119},
  123902 (2017).

\bibitem{Liu2015_OE_Ultra}
W.~Liu, \enquote{Ultra-directional super-scattering of homogenous spherical
  particles with radial anisotropy,} Opt. Express \textbf{23}, 14734 (2015).

\bibitem{Liu2012_ACSNANO}
W.~Liu, A.~E. Miroshnichenko, D.~N. Neshev, and Y.~S. Kivshar,
  \enquote{Broadband unidirectional scattering by magneto-electric core-shell
  nanoparticles,} ACS Nano \textbf{6}, 5489 (2012).

\bibitem{WANG_ACSNano_janus_2015}
H.~Wang, P.~Liu, Y.~Ke, Y.~Su, L.~Zhang, N.~Xu, S.~Deng, and H.~Chen,
  \enquote{Janus magneto\textendash{}electric nanosphere dimers exhibiting
  unidirectional visible light scattering and strong electromagnetic field
  enhancement,} ACS Nano \textbf{9}, 436 (2015).

\bibitem{ziolkowski_using_2017}
R.~W. Ziolkowski, \enquote{Using {Huygens} multipole arrays to realize
  unidirectional needle-like radiation,} Phys. Rev. X \textbf{7}, 031017
  (2017).

\bibitem{Alu2010_JN}
A.~Alu and N.~Engheta, \enquote{How does zero forward-scattering in
  magnetodielectric nanoparticles comply with the optical theorem?} J.
  Nanophotonics \textbf{4}, 041590 (2010).

\bibitem{Gomez-Medina2011_JN}
R.~Gomez-Medina, B.~Garcia-Camara, I.~Suarez-Lacalle, F.~Gonzalez, F.~Moreno,
  M.~Nieto-Vesperinas, and J.~J. Saenz, \enquote{Electric and magnetic dipolar
  response of germanium nanospheres: interference effects, scattering
  anisotropy, and optical forces,} J. Nanophotonics \textbf{5}, 053512 (2011).

\bibitem{Geffrin2012_NC}
J.~M. Geffrin, B.~Garcia-Camara, R.~Gomez-Medina, P.~Albella, L.~S.
  Froufe-Perez, C.~Eyraud, A.~Litman, R.~Vaillon, F.~Gonzalez,
  M.~Nieto-Vesperinas, J.~J. Saenz, and F.~Moreno, \enquote{Magnetic and
  electric coherence in forward- and back-scattered electromagnetic waves by a
  single dielectric subwavelength sphere,} Nat. Commun. \textbf{3}, 1171
  (2012).

\bibitem{Fu2013_NC}
Y.~H. Fu, A.~I. Kuznetsov, A.~E. Miroshnichenko, Y.~F. Yu, and B.~Lukyanchuk,
  \enquote{Directional visible light scattering by silicon nanoparticles,} Nat.
  Commun. \textbf{4}, 1527 (2013).

\bibitem{liu_scattering_2013-1}
W.~Liu, A.~E. Miroshnichenko, R.~F. Oulton, D.~N. Neshev, O.~Hess, and Y.~S.
  Kivshar, \enquote{Scattering of core-shell nanowires with the interference of
  electric and magnetic resonances,} Opt. Lett. \textbf{38}, 2621 (2013).

\bibitem{ZAMBRANA-PUYALTO2013Opt.Lett.}
X.~Zambrana-Puyalto, I.~Fernandez-Corbaton, M.~L. Juan, X.~Vidal, and
  G.~Molina-Terriza, \enquote{Duality symmetry and {{Kerker}} conditions,} Opt.
  Lett. \textbf{38}, 1857--1859 (2013).

\bibitem{liberal_superbackscattering_2015}
I.~Liberal, I.~Ederra, R.~Gonzalo, and R.~W. Ziolkowski,
  \enquote{Superbackscattering from single dielectric particles,} J. Opt.
  \textbf{17}, 072001 (2015).

\bibitem{PANIAGUA-DOMINGUEZ_NatCommun_generalized_2016-1}
R.~Paniagua-Dom{\'\i}nguez, Y.~F. Yu, A.~E. Miroshnichenko, L.~A. Krivitsky,
  Y.~H. Fu, V.~Valuckas, L.~Gonzaga, Y.~T. Toh, A.~Y.~S. Kay, B.~Luk'yanchuk,
  and A.~I. Kuznetsov, \enquote{Generalized {Brewster} effect in dielectric
  metasurfaces,} Nat. Commun. \textbf{7}, 10362 (2016).

\bibitem{LIU_Phys.Rev.A_superscattering_2017}
W.~Liu, \enquote{Superscattering pattern shaping for radially anisotropic
  nanowires,} Phys. Rev. A \textbf{96}, 023854 (2017).

\bibitem{wiecha_strongly_2017}
P.~R. Wiecha, A.~Cuche, A.~Arbouet, C.~Girard, G.~Colas~des Francs,
  A.~Lecestre, G.~Larrieu, F.~Fournel, V.~Larrey, T.~Baron, and V.~Paillard,
  \enquote{Strongly directional scattering from dielectric nanowires,} ACS
  Photonics \textbf{4}, 2036 (2017).

\bibitem{lee_reexamination_2017}
J.~Y. Lee, A.~E. Miroshnichenko, and R.-K. Lee, \enquote{Reexamination of
  {Kerker}'s conditions by means of the phase diagram,} Phys. Rev. A
  \textbf{96}, 043846 (2017).

\bibitem{GOUESBET_generalized_2011}
G.~Gouesbet and G.~Gr{\'e}han, \emph{Generalized {{Lorenz}}-{{Mie Theories}}}
  ({Springer Science \& Business Media}, 2011).

\bibitem{WANG_Opt.Lett._optimized_2014}
N.~Wang, W.~Lu, J.~Ng, and Z.~Lin, \enquote{Optimized optical ``tractor beam''
  for core\textendash{}shell nanoparticles,} Opt. Lett. \textbf{39}, 2399
  (2014).

\bibitem{WOZNIAK_LaserPhotonicsRev._selective_2015}
P.~Wo{\'z}niak, P.~Banzer, and G.~Leuchs, \enquote{Selective switching of
  individual multipole resonances in single dielectric nanoparticles,} Laser
  Photonics Rev. \textbf{9}, 231 (2015).

\bibitem{NEUGEBAUER_NatCommun_polarizationcontrolled_2016}
M.~Neugebauer, P.~Wo{\'z}niak, A.~Bag, G.~Leuchs, and P.~Banzer,
  \enquote{Polarization-controlled directional scattering for nanoscopic
  position sensing,} Nat. Commun. \textbf{7}, 11286 (2016).

\bibitem{WEI_Opt.Lett.OL_adding_2017}
L.~Wei, N.~Bhattacharya, and H.~P. Urbach, \enquote{Adding a spin to
  {{Kerker}}'s condition: angular tuning of directional scattering with
  designed excitation,} Opt. Lett. \textbf{42}, 1776 (2017).

\bibitem{XI_Phys.Rev.Lett._magnetic_2017}
Z.~Xi and H.~P. Urbach, \enquote{Magnetic dipole scattering from metallic
  nanowire for ultrasensitive deflection sensing,} Phys. Rev. Lett.
  \textbf{119}, 053902 (2017).

\bibitem{LIU_Opt.ExpressOE_optical_2017}
H.~Liu, M.~Panmai, Y.~Peng, and S.~Lan, \enquote{Optical pulling and pushing
  forces exerted on silicon nanospheres with strong coherent interaction
  between electric and magnetic resonances,} Opt. Express \textbf{25}, 12357
  (2017).

\bibitem{MELIK-GAYKAZYAN_2017_ACSPhotonics_Selective}
E.~V. Melik-Gaykazyan, S.~S. Kruk, R.~Camacho-Morales, L.~Xu, M.~Rahmani,
  K.~Zangeneh~Kamali, A.~Lamprianidis, A.~E. Miroshnichenko, A.~A. Fedyanin,
  D.~N. Neshev, and Y.~S. Kivshar, \enquote{Selective third-harmonic generation
  by structured light in {{Mie}}-resonant nanoparticles,} ACS Photonics, doi:
  10.1021/acsphotonics.7b01277  (2017).

\bibitem{Novotny2012_book}
L.~Novotny and B.~Hecht, \emph{Principles of {Nano}-{Optics}} (Cambridge
  University, Cambridge, 2012).

\bibitem{GROEP2013Opt.Express}
J.~van~de Groep and A.~Polman, \enquote{Designing dielectric resonators on
  substrates: combining magnetic and electric resonances,} Opt. Express
  \textbf{21}, 26285 (2013).

\bibitem{PORS_Opt.Express_unidirectional_2015}
A.~Pors, S.~K. Andersen, and S.~I. Bozhevolnyi, \enquote{Unidirectional
  scattering by nanoparticles near substrates: generalized {{Kerker}}
  conditions,} Opt. Express \textbf{23}, 28808 (2015).

\bibitem{SINEV2016Laser&PhotonicsReviews}
I.~Sinev, I.~Iorsh, A.~Bogdanov, D.~Permyakov, F.~Komissarenko, I.~Mukhin,
  A.~Samusev, V.~Valuckas, A.~I. Kuznetsov, B.~S. Luk'yanchuk, A.~E.
  Miroshnichenko, and Y.~S. Kivshar, \enquote{Polarization control over
  electric and magnetic dipole resonances of dielectric nanoparticles on
  metallic films,} Laser Photonics Rev. \textbf{10}, 799 (2016).

\bibitem{HOLSTEEN2017Science}
A.~L. Holsteen, S.~Raza, P.~Fan, P.~G. Kik, and M.~L. Brongersma,
  \enquote{Purcell effect for active tuning of light scattering from
  semiconductor optical antennas,} Science \textbf{358}, 1407 (2017).

\bibitem{Hancu2013_NL}
I.~M. Hancu, A.~G. Curto, M.~Castro-L¨®pez, M.~Kuttge, and N.~F. van Hulst,
  \enquote{Multipolar interference for directed light emission,} Nano Lett.
  \textbf{14}, 166 (2013).

\bibitem{LI_NanoLett._alldielectric_2016}
J.~Li, N.~Verellen, D.~Vercruysse, T.~Bearda, L.~Lagae, and P.~{Van Dorpe},
  \enquote{All-dielectric antenna wavelength router with bidirectional
  scattering of visible light,} Nano Lett. \textbf{16}, 4396 (2016).

\bibitem{COENEN_NatCommun_directional_2014}
T.~Coenen, F.~Bernal~Arango, A.~Femius~Koenderink, and A.~Polman,
  \enquote{Directional emission from a single plasmonic scatterer,} Nat.
  Commun. \textbf{5}, 3250 (2014).

\bibitem{WANG2016Phys.Rev.Lett.}
Z.~Wang, K.~Yao, M.~Chen, H.~Chen, and Y.~Liu, \enquote{Manipulating
  {{Smith}}-{{Purcell}} emission with {{Babinet}} metasurfaces,} Phys. Rev.
  Lett. \textbf{117}, 157401 (2016).

\bibitem{GURUNARAYANAN2017NanoLett.}
S.~P. Gurunarayanan, N.~Verellen, V.~S. Zharinov, F.~James~Shirley, V.~V.
  Moshchalkov, M.~Heyns, J.~{Van de Vondel}, I.~P. Radu, and P.~Van~Dorpe,
  \enquote{Electrically driven unidirectional optical nanoantennas,} Nano Lett.
  \textbf{17}, 7433 (2017).

\bibitem{Rolly2013_arxiv}
B.~Rolly, R.~Abdeddaim, J.-M. Geffrin, B.~Stout, and N.~Bonod,
  \enquote{Controllable emission of a dipolar source coupled with a
  magneto-dielectric resonant subwavelength scatterer,} Sci. Rep. \textbf{3},
  3063 (2013).

\bibitem{Krasnok2013_arxiv}
A.~E. Krasnok, C.~R. Simovski, P.~A. Belov, and Y.~S. Kivshar,
  \enquote{Superdirective dielectric nanoantennas,} Nanoscale \textbf{6}, 7354
  (2014).

\bibitem{VERCRUYSSE_ACSNano_directional_2014}
D.~Vercruysse, X.~Zheng, Y.~Sonnefraud, N.~Verellen, G.~{Di Martino}, L.~Lagae,
  G.~A.~E. Vandenbosch, V.~V. Moshchalkov, S.~A. Maier, and P.~{Van Dorpe},
  \enquote{Directional fluorescence emission by individual {{V}}-antennas
  explained by mode expansion,} ACS Nano \textbf{8}, 8232 (2014).

\bibitem{NARAGHI_Opt.Lett.OL_directional_2015}
R.~R. Naraghi, S.~Sukhov, and A.~Dogariu, \enquote{Directional control of
  scattering by all-dielectric core-shell spheres,} Opt. Lett. \textbf{40}, 585
  (2015).

\bibitem{DEABAJO_Rev.Mod.Phys._colloquium_2007}
F.~G. De~Abajo, \enquote{Colloquium: {{Light}} scattering by particle and hole
  arrays,} Rev. Mod. Phys. \textbf{79}, 1267 (2007).

\bibitem{KIHM_Nat.Commun._bethehole_2011}
H.~W. Kihm, S.~M. Koo, Q.~H. Kim, K.~Bao, J.~E. Kihm, W.~S. Bak, S.~H. Eah,
  C.~Lienau, H.~Kim, P.~Nordlander, N.~J. Halas, N.~K. Park, and D.-S. Kim,
  \enquote{Bethe-hole polarization analyser for the magnetic vector of light,}
  Nat. Commun. \textbf{2}, 451 (2011).

\bibitem{GROSJEAN_NanoLett._diabolo_2011}
T.~Grosjean, M.~Mivelle, F.~I. Baida, G.~W. Burr, and U.~C. Fischer,
  \enquote{Diabolo nanoantenna for enhancing and confining the magnetic optical
  field,} Nano Lett. \textbf{11}, 1009 (2011).

\bibitem{COENEN_ACSNano_optical_2014}
T.~Coenen and A.~Polman, \enquote{Optical properties of single plasmonic holes
  probed with local electron beam excitation,} ACS Nano \textbf{8}, 7350
  (2014).

\bibitem{ROTENBERG_NatPhoton_mapping_2014}
N.~Rotenberg and L.~Kuipers, \enquote{Mapping nanoscale light fields,} Nat.
  Photonics \textbf{8}, 919 (2014).

\bibitem{YAN2017LightSci.Appl.}
J.~Yan, C.~Ma, P.~Liu, C.~Wang, and G.~Yang, \enquote{Generating scattering
  dark states through the {{Fano}} interference between excitons and an
  individual silicon nanogroove,} Light Sci. Appl. \textbf{6}, e16197 (2017).

\bibitem{HUIDOBRO_Phys.Rev.X_magnetic_2014}
P.~A. Huidobro, X.~Shen, J.~Cuerda, E.~Moreno, L.~Martin-Moreno,
  J.~Garcia-Vidal, F.\, T.~J. Cui, and B.~Pendry, J.\, \enquote{Magnetic
  localized surface plasmons,} Phys. Rev. X \textbf{4}, 021003 (2014).

\bibitem{WU2018ArXiv180104040Phys.}
H.-W. Wu, H.-J. Chen, H.-F. Xu, Y.~Zhou, D.-X. Qi, and Y.-K. Wang,
  \enquote{Tunable multiband directional electromagnetic scattering from spoof
  {{Mie}} resonant structure,} arXiv180104040  (2018).

\bibitem{SAUTTER2015ACSNano}
J.~Sautter, I.~Staude, M.~Decker, E.~Rusak, D.~N. Neshev, I.~Brener, and Y.~S.
  Kivshar, \enquote{Active tuning of all-dielectric metasurfaces,} ACS Nano
  \textbf{9}, 4308 (2015).

\bibitem{ALAEE_Opt.Lett._phasechange_2016}
R.~Alaee, M.~Albooyeh, S.~Tretyakov, and C.~Rockstuhl, \enquote{Phase-change
  material-based nanoantennas with tunable radiation patterns,} Opt. Lett.
  \textbf{41}, 4099 (2016).

\bibitem{WUTTIG_NatPhoton_phasechange_2017}
M.~Wuttig, H.~Bhaskaran, and T.~Taubner, \enquote{Phase-change materials for
  non-volatile photonic applications,} Nat. Photonics \textbf{11}, 465 (2017).

\bibitem{RUTCKAIA2017NanoLett.}
V.~Rutckaia, F.~Heyroth, A.~Novikov, M.~Shaleev, M.~Petrov, and J.~Schilling,
  \enquote{Quantum dot emission driven by mie resonances in silicon
  nanostructures,} Nano Lett. \textbf{17}, 6886 (2017).

\bibitem{BARANOV2017Laser&PhotonicsReviews}
D.~G. Baranov, R.~S. Savelev, S.~V. Li, A.~E. Krasnok, and A.~Al{\`u},
  \enquote{Modifying magnetic dipole spontaneous emission with nanophotonic
  structures,} Laser Photonics Rev. \textbf{11}, 1600268 (2017).

\bibitem{KRUK_ACSPhotonics_spinpolarized_2014}
S.~S. Kruk, M.~Decker, I.~Staude, S.~Schlecht, M.~Greppmair, D.~N. Neshev, and
  Y.~S. Kivshar, \enquote{Spin-polarized photon emission by resonant multipolar
  nanoantennas,} ACS Photonics \textbf{1}, 1218 (2014).

\bibitem{GARCIA-ETXARRI2017ACSPhotonics}
A.~Garcia-Etxarri, \enquote{Optical polarization {M{\"o}bius} strips on
  all-dielectric optical scatterers,} ACS Photonics \textbf{4}, 1159 (2017).

\bibitem{NOVITSKY_Phys.Rev.Lett._materialindependent_2012}
A.~Novitsky, C.-W. Qiu, and A.~Lavrinenko, \enquote{Material-independent and
  size-independent tractor beams for dipole objects,} Phys. Rev. Lett.
  \textbf{109}, 023902 (2012).

\bibitem{GAO_LightSci.Appl._optical_2017}
D.~Gao, W.~Ding, M.~Nieto-Vesperinas, X.~Ding, M.~Rahman, T.~Zhang, C.~Lim, and
  C.-W. Qiu, \enquote{Optical manipulation from the microscale to the
  nanoscale: fundamentals, advances and prospects,} Light Sci. Appl.
  \textbf{6}, e17039 (2017).

\bibitem{ROTENBERG_Phys.Rev.Lett._plasmon_2012}
N.~Rotenberg, M.~Spasenovi{\'c}, T.~L. Krijger, B.~Le~Feber, F.~G. {de Abajo},
  and L.~Kuipers, \enquote{Plasmon scattering from single subwavelength holes,}
  Phys. Rev. Lett. \textbf{108}, 127402 (2012).

\bibitem{LEE_Phys.Rev.Lett._role_2012}
S.-Y. Lee, I.-M. Lee, J.~Park, S.~Oh, W.~Lee, K.-Y. Kim, and B.~Lee,
  \enquote{Role of magnetic induction currents in nanoslit excitation of
  surface plasmon polaritons,} Phys. Rev. Lett. \textbf{108}, 213907 (2012).

\bibitem{RODRIGUEZ-FORTUNO_Science_nearfield_2013}
F.~J. Rodr{\'\i}guez-Fortu{\~n}o, G.~Marino, P.~Ginzburg, D.~O'Connor,
  A.~Mart{\'\i}nez, G.~A. Wurtz, and A.~V. Zayats, \enquote{Near-field
  interference for the unidirectional excitation of electromagnetic guided
  modes,} Science \textbf{340}, 328 (2013).

\bibitem{EVLYUKHIN_Phys.Rev.B_resonant_2015}
A.~B. Evlyukhin and S.~I. Bozhevolnyi, \enquote{Resonant unidirectional and
  elastic scattering of surface plasmon polaritons by high refractive index
  dielectric nanoparticles,} Phys. Rev. B \textbf{92}, 245419 (2015).

\bibitem{SINEV_Laser&PhotonicsReviews_chirality_2017}
I.~S. Sinev, A.~A. Bogdanov, F.~E. Komissarenko, K.~S. Frizyuk, M.~I. Petrov,
  I.~S. Mukhin, S.~V. Makarov, A.~K. Samusev, A.~V. Lavrinenko, and I.~V.
  Iorsh, \enquote{Chirality driven by magnetic dipole response for
  demultiplexing of surface waves,} Laser Photonics Rev. \textbf{11}, 1700168
  (2017).

\bibitem{PICARDI_Phys.Rev.B_unidirectional_2017}
M.~F. Picardi, A.~Manjavacas, A.~V. Zayats, and F.~J.
  Rodr{\'\i}guez-Fortu{\~n}o, \enquote{Unidirectional evanescent-wave coupling
  from circularly polarized electric and magnetic dipoles: {{An}} angular
  spectrum approach,} Phys. Rev. B \textbf{95}, 245416 (2017).

\bibitem{PICARDI_ArXiv170802494Phys._janus_2017}
M.~F. Picardi, A.~V. Zayats, and F.~J. Rodr{\'\i}guez-Fortu{\~n}o,
  \enquote{Janus and huygens dipoles: near-field directionality beyond
  spin-momentum locking,} Phys. Rev. Lett. \textbf{120}, 117402 (2018), preprint: arXiv170802494.

\bibitem{BLIOKH_Nat.Photonics_spinorbit_2015}
K.~Y. Bliokh, F.~J. Rodr{\'\i}guez-Fortu{\~n}o, F.~Nori, and A.~V. Zayats,
  \enquote{Spin-orbit interactions of light,} Nat. Photonics \textbf{9}, 796
  (2015).

\bibitem{VANMECHELEN_Optica_universal_2016}
T.~{Van Mechelen} and Z.~Jacob, \enquote{Universal spin-momentum locking of
  evanescent waves,} Optica \textbf{3}, 118 (2016).

\bibitem{QUINTEN_optical_2010}
M.~Quinten, \emph{Optical Properties of Nanoparticle Systems: {{Mie}} and
  Beyond} ({John Wiley \& Sons}, 2010).

\bibitem{Merchiers2007_PRA}
O.~Merchiers, F.~Moreno, F.~Gonzalez, and J.~M. Saiz, \enquote{Light scattering
  by an ensemble of interacting dipolar particles with both electric and
  magnetic polarizabilities,} Phys. Rev. A \textbf{76}, 043834 (2007).

\bibitem{YAN_Nat.Commun._magnetically_2015}
J.~H. Yan, P.~Liu, Z.~Y. Lin, H.~Wang, H.~J. Chen, C.~X. Wang, and G.~W. Yang,
  \enquote{Magnetically induced forward scattering at visible wavelengths in
  silicon nanosphere oligomers,} Nat. Commun. \textbf{6}, 7042 (2015).

\bibitem{VANDEGROEP_Optica_direct_2016}
J.~{van de Groep}, T.~Coenen, S.~A. Mann, and A.~Polman, \enquote{Direct
  imaging of hybridized eigenmodes in coupled silicon nanoparticles,} Optica
  \textbf{3}, 93 (2016).

\bibitem{SHIBANUMA_ACSPhotonics_experimental_2017}
T.~Shibanuma, T.~Matsui, T.~Roschuk, J.~Wojcik, P.~Mascher, P.~Albella, and
  S.~A. Maier, \enquote{Experimental demonstration of tunable directional
  scattering of visible light from all-dielectric asymmetric dimers,} ACS
  Photonics \textbf{4}, 489 (2017).

\bibitem{BARREDA_Nat.Commun._electromagnetic_2017}
{\'A}.~I. Barreda, H.~Saleh, A.~Litman, F.~Gonz{\'a}lez, J.-M. Geffrin, and
  F.~Moreno, \enquote{Electromagnetic polarization-controlled perfect switching
  effect with high-refractive-index dimers and the beam-splitter
  configuration,} Nat. Commun. \textbf{8}, 1038 (2017).

\bibitem{LU2015Laser&PhotonicsReviews}
G.~Lu, Y.~Wang, R.~Y. Chou, H.~Shen, Y.~He, Y.~Cheng, and Q.~Gong,
  \enquote{Directional side scattering of light by a single plasmonic trimer,}
  Laser Photonics Rev. \textbf{9}, 530 (2015).

\bibitem{BANZER_Nat.Commun._chiral_2016}
P.~Banzer, P.~Wo{\'z}niak, U.~Mick, I.~D. Leon, and R.~W. Boyd, \enquote{Chiral
  optical response of planar and symmetric nanotrimers enabled by
  heteromaterial selection,} Nat. Commun. \textbf{7}, 13117 (2016).

\bibitem{YAO_ACSPhotonics_controlling_2016}
K.~Yao and Y.~Liu, \enquote{Controlling electric and magnetic resonances for
  ultracompact nanoantennas with tunable directionality,} ACS Photonics
  \textbf{3}, 953 (2016).

\bibitem{HOPKINS_ACSPhotonics_interplay_2015}
B.~Hopkins, D.~S. Filonov, A.~E. Miroshnichenko, F.~Monticone, A.~Al{\`u}, and
  Y.~S. Kivshar, \enquote{Interplay of magnetic responses in all-dielectric
  oligomers to realize magnetic {Fano} resonances,} ACS Photonics \textbf{2},
  724 (2015).

\bibitem{Miroshnichenko2012_NL6459}
A.~E. Miroshnichenko and Y.~S. Kivshar, \enquote{Fano resonances in
  all-dielectric oligomers,} Nano Lett. \textbf{12}, 6459 (2012).

\bibitem{CAO_NanoLett._optical_2011}
L.~Cao, P.~Fan, and M.~L. Brongersma, \enquote{Optical coupling of
  deep-subwavelength semiconductor nanowires,} Nano Lett. \textbf{11}, 1463
  (2011).

\bibitem{LIMONOV_NatPhoton_fano_2017}
M.~F. Limonov, M.~V. Rybin, A.~N. Poddubny, and Y.~S. Kivshar, \enquote{Fano
  resonances in photonics,} Nat. Photonics \textbf{11}, 543 (2017).

\bibitem{LIU_ArXivPrepr.ArXiv170406049_scattering_2017}
W.~Liu and A.~E. Miroshnichenko, \enquote{Scattering invisibility with
  free-space field enhancement of all-dielectric nanoparticles,} Laser
  Photonics Rev. \textbf{11}, 201700103 (2017).

\bibitem{HAUS_waves_1984}
H.~A. Haus, \emph{Waves and Fields in Optoelectronics} (Prentice-Hall, 1984).

\bibitem{BYELOBROV_IEEEAntennasPropag.Mag._periodicity_2015}
V.~O. Byelobrov, T.~L. Zinenko, K.~Kobayashi, and A.~I. Nosich,
  \enquote{Periodicity matters: grating or lattice resonances in the scattering
  by sparse arrays of subwavelength strips and wires.} IEEE Antennas Propag.
  Mag. \textbf{57}, 34 (2015).

\bibitem{Liu2012_PRB}
W.~Liu, A.~E. Miroshnichenko, D.~N. Neshev, and Y.~S. Kivshar,
  \enquote{Polarization-independent {Fano} resonances in arrays of core-shell
  nanoparticles,} Phys. Rev. B \textbf{86}, 081407(\textbf{R}) (2012).

\bibitem{Joannopoulos2008_book}
J.~D. Joannopoulos, \emph{Photonic {{Crystals}} : {{Molding}} the {{Flow}} of
  {{Light}}} (Princeton University, Princeton, 2008).

\bibitem{decker_high-efficiency_2015}
M.~Decker, I.~Staude, M.~Falkner, J.~Dominguez, D.~N. Neshev, I.~Brener,
  T.~Pertsch, and Y.~S. Kivshar, \enquote{High-efficiency dielectric {Huygens}
  surfaces,} Adv. Opt. Mater. \textbf{3}, 813 (2015).

\bibitem{KIM_Phys.Rev.X_optical_2014}
M.~Kim, H.~Wong, Alex~M.\, and G.~V. Eleftheriades, \enquote{Optical
  {Huygens¡¯} metasurfaces with independent control of the magnitude and phase
  of the local reflection coefficients,} Phys. Rev. X \textbf{4}, 041042
  (2014).

\bibitem{WANG_Appl.Phys.Lett._generation_2014}
F.~Wang, Q.-H. Wei, and H.~Htoon, \enquote{Generation of steep phase anisotropy
  with zero-backscattering by arrays of coupled dielectric nano-resonators,}
  Appl. Phys. Lett. \textbf{105}, 121112 (2014).

\bibitem{LIU_NanoLett._huygens_2017-1}
S.~Liu, A.~Vaskin, S.~Campione, O.~Wolf, M.~B. Sinclair, J.~Reno, G.~A. Keeler,
  I.~Staude, and I.~Brener, \enquote{Huygens¡¯ metasurfaces enabled by magnetic
  dipole resonance tuning in split dielectric nanoresonators,} Nano Lett.
  \textbf{17}, 4297 (2017).

\bibitem{BABICHEVA_Laser&PhotonicsReviews_resonant}
V.~E. Babicheva and A.~B. Evlyukhin, \enquote{Resonant lattice {{Kerker}}
  effect in metasurfaces with electric and magnetic optical responses,} Laser
  Photonics Rev. \textbf{11}, 1700132 (2017).

\bibitem{VASILANTONAKIS_ArXiv171101430Phys._designing_2017}
N.~Vasilantonakis, J.~Scheuer, and A.~Boag, \enquote{Designing
  high-transmission and wide angle all-dielectric flat metasurfaces at telecom
  wavelengths,} arXiv171101430  (2017).

\bibitem{CHENKE_2017_AdvancedMaterials_Reconfigurable}
K.~Chen, Y.~Feng, F.~Monticone, J.~Zhao, B.~Zhu, T.~Jiang, L.~Zhang, Y.~Kim,
  X.~Ding, S.~Zhang \emph{et~al.}, \enquote{A reconfigurable active huygens
  metalens,} Adv. Materials \textbf{29}, 1606422 (2017).

\bibitem{KRUK_APLPhotonics_invited_2016}
S.~Kruk, B.~Hopkins, I.~I. Kravchenko, A.~Miroshnichenko, D.~N. Neshev, and
  Y.~S. Kivshar, \enquote{Broadband highly efficient dielectric metadevices for
  polarization control,} APL Photonics \textbf{1}, 030801 (2016).

\bibitem{Spinelli2012_NC}
P.~Spinelli, M.~A. Verschuuren, and A.~Polman, \enquote{Broadband
  omnidirectional antireflection coating based on subwavelength surface {Mie}
  resonators,} Nat. Commun. \textbf{3}, 692 (2012).

\bibitem{RODRIGUEZ_Phys.Rev.Lett._breaking_2014}
S.~R.~K. Rodriguez, F.~B. Arango, T.~P. Steinbusch, M.~A. Verschuuren, A.~F.
  Koenderink, and J.~G. Rivas, \enquote{Breaking the symmetry of
  forward-backward light emission with localized and collective magnetoelectric
  resonances in arrays of pyramid-shaped aluminum nanoparticles,} Phys. Rev.
  Lett. \textbf{113}, 247401 (2014).

\bibitem{CAMPIONE_Opt.ExpressOE_tailoring_2015}
S.~Campione, L.~I. Basilio, L.~K. Warne, and M.~B. Sinclair, \enquote{Tailoring
  dielectric resonator geometries for directional scattering and {{Huygens}}'
  metasurfaces,} Opt. Express \textbf{23}, 2293 (2015).

\bibitem{YANG_Phys.Rev.B_multimode_2017-1}
Y.~Yang, A.~E. Miroshnichenko, S.~V. Kostinski, M.~Odit, P.~Kapitanova, M.~Qiu,
  and Y.~S. Kivshar, \enquote{Multimode directionality in all-dielectric
  metasurfaces,} Phys. Rev. B \textbf{95}, 165426 (2017).

\bibitem{liu_beam_2017}
W.~Liu and A.~E. Miroshnichenko, \enquote{Beam steering with dielectric
  metalattices,} ACS Photonics, doi:10.1021/acsphotonics.7b01217  (2017).

\bibitem{YANG_NatCommun_alldielectric_2014}
Y.~Yang, I.~I. Kravchenko, D.~P. Briggs, and J.~Valentine,
  \enquote{All-dielectric metasurface analogue of electromagnetically induced
  transparency,} Nat. Commun. \textbf{5}, 5753 (2014).

\bibitem{WU_NatCommun_spectrally_2014}
C.~Wu, N.~Arju, G.~Kelp, J.~A. Fan, J.~Dominguez, E.~Gonzales, E.~Tutuc,
  I.~Brener, and G.~Shvets, \enquote{Spectrally selective chiral silicon
  metasurfaces based on infrared {{Fano}} resonances,} Nat. Commun. \textbf{5},
  3892 (2014).

\bibitem{ZHAO_Opt.ExpressOE_dipole_2014}
W.~Zhao, D.~Ju, Y.~Jiang, and Q.~Zhan, \enquote{Dipole and quadrupole trapped
  modes within bi-periodic siliconparticle array realizing three-channel
  refractive sensing,} Opt. Express \textbf{22}, 31277 (2014).

\bibitem{GHENUCHE_Phys.Rev.Lett._optical_2012}
P.~Ghenuche, G.~Vincent, M.~Laroche, N.~Bardou, R.~Ha{\"\i}dar, J.-L. Pelouard,
  and S.~Collin, \enquote{Optical extinction in a single layer of nanorods,}
  Phys. Rev. Lett. \textbf{109}, 143903 (2012).

\bibitem{DU_Phys.Rev.Lett._nearly_2013}
J.~Du, Z.~Lin, S.~T. Chui, G.~Dong, and W.~Zhang, \enquote{Nearly total
  omnidirectional reflection by a single layer of nanorods,} Phys. Rev. Lett.
  \textbf{110}, 163902 (2013).

\bibitem{SLOVICK2013Phys.Rev.B}
B.~Slovick, Z.~G. Yu, M.~Berding, and S.~Krishnamurthy, \enquote{Perfect
  dielectric-metamaterial reflector,} Phys. Rev. B \textbf{88}, 165116 (2013).

\bibitem{LIU_OpticaOPTICA_optical_2014}
S.~Liu, M.~B. Sinclair, T.~S. Mahony, Y.~C. Jun, S.~Campione, J.~Ginn, D.~A.
  Bender, J.~R. Wendt, J.~F. Ihlefeld, P.~G. Clem, J.~B. Wright, and I.~Brener,
  \enquote{Optical magnetic mirrors without metals,} Optica \textbf{1}, 250
  (2014).

\bibitem{MOITRA_ACSPhotonics_largescale_2015}
P.~Moitra, B.~A. Slovick, W.~Li, I.~I. Kravchencko, D.~P. Briggs,
  S.~Krishnamurthy, and J.~Valentine, \enquote{Large-scale all-dielectric
  metamaterial perfect reflectors,} ACS Photonics \textbf{2}, 692 (2015).

\bibitem{shalaev_high-efficiency_2015}
M.~I. Shalaev, J.~Sun, A.~Tsukernik, A.~Pandey, K.~Nikolskiy, and N.~M.
  Litchinitser, \enquote{High-efficiency all-dielectric metasurfaces for
  ultracompact beam manipulation in transmission mode,} Nano Lett. \textbf{15},
  6261 (2015).

\bibitem{LI_2017_ACSNano_Dielectric}
Z.~Li, I.~Kim, L.~Zhang, M.~Q. Mehmood, M.~S. Anwar, M.~Saleem, D.~Lee, K.~T.
  Nam, S.~Zhang, B.~Luk'yanchuk, Y.~Wang, G.~Zheng, J.~Rho, and C.-W. Qiu,
  \enquote{Dielectric meta-holograms enabled with dual magnetic resonances in
  visible light,} ACS Nano \textbf{11}, 9382 (2017).

\bibitem{DU_Phys.Rev.Lett._optical_2011}
J.~Du, Z.~Lin, S.~T. Chui, W.~Lu, H.~Li, A.~Wu, Z.~Sheng, J.~Zi, X.~Wang,
  S.~Zou, and F.~Gan, \enquote{Optical beam steering based on the symmetry of
  resonant modes of nanoparticles,} Phys. Rev. Lett. \textbf{106}, 203903
  (2011).

\bibitem{WU_NanoLett._experimental_2015}
A.~Wu, H.~Li, J.~Du, X.~Ni, Z.~Ye, Y.~Wang, Z.~Sheng, S.~Zou, F.~Gan, X.~Zhang,
  and X.~Wang, \enquote{Experimental demonstration of in-plane negative-angle
  refraction with an array of silicon nanoposts,} Nano Lett. \textbf{15}, 2055
  (2015).

\bibitem{RADI_Phys.Rev.Lett._metagratings_2017}
Y.~Ra'di, D.~L. Sounas, and A.~Al{\`u}, \enquote{Metagratings: beyond the
  limits of graded metasurfaces for wave front control,} Phys. Rev. Lett.
  \textbf{119}, 067404 (2017).

\bibitem{CHALABI_Phys.Rev.B_efficient_2017}
H.~Chalabi, Y.~Ra'di, D.~L. Sounas, and A.~Al{\`u}, \enquote{Efficient
  anomalous reflection through near-field interactions in metasurfaces,} Phys.
  Rev. B \textbf{96}, 075432 (2017).

\bibitem{PANIAGUA-DOMINGUEZ_ArXiv170500895Phys._metalens_2017}
R.~Paniagua-Dom{\'\i}nguez, Y.~F. Yu, E.~Khaidarov, S.~Choi, V.~Leong, R.~M.
  Bakker, X.~Liang, Y.~H. Fu, V.~Valuckas, L.~A. Krivitsky, and A.~I.
  Kuznetsov, \enquote{A metalens with a near-unity numerical aperture,} Nano
  Lett., doi:10.1021/acs.nanolett.8b00368  (2018).

\bibitem{KHAIDAROV_NanoLett._asymmetric_2017}
E.~Khaidarov, H.~Hao, R.~Paniagua-Dom{\'\i}nguez, Y.~F. Yu, Y.~H. Fu,
  V.~Valuckas, S.~L.~K. Yap, Y.~T. Toh, J.~S.~K. Ng, and A.~I. Kuznetsov,
  \enquote{Asymmetric nanoantennas for ultrahigh angle broadband visible light
  bending,} Nano Lett. \textbf{17}, 6267 (2017).

\bibitem{WATTS_Adv.Mater._metamaterial_2012}
C.~M. Watts, X.~Liu, and W.~J. Padilla, \enquote{Metamaterial electromagnetic
  wave absorbers,} Adv. Mater. \textbf{24}, OP98 (2012).

\bibitem{RADI_Phys.Rev.Appl._thin_2015}
Y.~Ra'Di, C.~R. Simovski, and S.~A. Tretyakov, \enquote{Thin perfect absorbers
  for electromagnetic waves: Theory, design, and realizations,} Phys. Rev.
  Appl. \textbf{3}, 037001 (2015).

\bibitem{BARANOV2017Nat.Rev.Mater.}
D.~G. Baranov, A.~Krasnok, T.~Shegai, A.~Al{\`u}, and Y.~Chong,
  \enquote{Coherent perfect absorbers: linear control of light with light,}
  Nat. Rev. Mater. \textbf{2}, 17064 (2017).

\bibitem{LAROCHE_ArXivPrepr.Physics0606119_controlling_2006}
M.~Laroche, R.~G{\'o}mez-Medina, and J.~J. Saenz, \enquote{Controlling
  absorption resonances from sub-wavelength cylinder arrays,} arXiv0606119
  (2006).

\bibitem{RADI_ACSPhotonics_full_2015}
Y.~Ra'di, V.~S. Asadchy, S.~U. Kosulnikov, M.~M. Omelyanovich, D.~Morits, A.~V.
  Osipov, C.~R. Simovski, and S.~A. Tretyakov, \enquote{Full light absorption
  in single arrays of spherical nanoparticles,} ACS Photonics \textbf{2}, 653
  (2015).

\bibitem{ODEBOLANK_NanoLett._largescale_2017}
N.~Odebo~L{\"a}nk, R.~Verre, P.~Johansson, and M.~K{\"a}ll,
  \enquote{Large-scale silicon nanophotonic metasurfaces with polarization
  independent near-perfect absorption,} Nano Lett. \textbf{17}, 3054 (2017).

\bibitem{ALAEE_ArXiv171108203Phys._theory_2017}
R.~Alaee, M.~Albooyeh, and C.~Rockstuhl, \enquote{Theory of metasurface based
  perfect absorbers,} J. Phys. D \textbf{50}, 503002 (2017).

\bibitem{YANG_2018_ACSPhotonics}
C.-Y. Yang, J.-H. Yang, Z.-Y. Yang, Z.-X. Zhou, M.-G. Sun, V.~E. Babicheva, and
  K.-P. Chen, \enquote{Nonradiating {{Silicon Nanoantenna Metasurfaces}} as
  {{Narrowband Absorbers}},} ACS Photonics, doi:10.1021/acsphotonics.7b01186
  (2018).

\bibitem{MOREAU_Nature_controlledreflectance_2012}
A.~Moreau, C.~Cirac{\`\i}, J.~J. Mock, R.~T. Hill, Q.~Wang, B.~J. Wiley,
  A.~Chilkoti, and D.~R. Smith, \enquote{Controlled-reflectance surfaces with
  film-coupled colloidal nanoantennas,} Nature \textbf{492}, 86 (2012).

\bibitem{AKSELROD_Adv.Mater._largearea_2015}
G.~M. Akselrod, J.~Huang, T.~B. Hoang, P.~T. Bowen, L.~Su, D.~R. Smith, and
  M.~H. Mikkelsen, \enquote{Large-area metasurface perfect absorbers from
  visible to near-infrared,} Adv. Mater. \textbf{27}, 8028 (2015).

\bibitem{SUN_Appl.Phys.Lett._broadband_2008}
C.-H. Sun, P.~Jiang, and B.~Jiang, \enquote{Broadband moth-eye antireflection
  coatings on silicon,} Appl. Phys. Lett. \textbf{92}, 061112 (2008).

\bibitem{CHEN_Phys.Rev.Lett._antireflection_2010}
H.-T. Chen, J.~Zhou, J.~F. O'Hara, F.~Chen, A.~K. Azad, and A.~J. Taylor,
  \enquote{Antireflection coating using metamaterials and identification of its
  mechanism,} Phys. Rev. Lett. \textbf{105}, 073901 (2010).

\bibitem{CHANG-HASNAIN_Adv.Opt.Photon.AOP_highcontrast_2012}
C.~J. Chang-Hasnain and W.~Yang, \enquote{High-contrast gratings for integrated
  optoelectronics,} Adv. Opt. Photon. \textbf{4}, 379 (2012).

\bibitem{COLLIN_Rep.Prog.Phys._nanostructure_2014}
S.~Collin, \enquote{Nanostructure arrays in free-space: optical properties and
  applications,} Rep. Prog. Phys. \textbf{77}, 126402 (2014).

\bibitem{ARBABI_Nat.Nanotechnol._dielectric_2015}
A.~Arbabi, Y.~Horie, M.~Bagheri, and A.~Faraon, \enquote{Dielectric
  metasurfaces for complete control of phase and polarization with
  subwavelength spatial resolution and high transmission,} Nat. Nanotechnol.
  \textbf{10}, 937 (2015).

\bibitem{ZHAN_ACSPhotonics_lowcontrast_2016}
A.~Zhan, S.~Colburn, R.~Trivedi, T.~K. Fryett, C.~M. Dodson, and A.~Majumdar,
  \enquote{Low-contrast dielectric metasurface optics,} ACS Photonics
  \textbf{3}, 209 (2016).

\bibitem{lalanne_metalenses_2017}
P.~Lalanne and P.~Chavel, \enquote{Metalenses at visible wavelengths: past,
  present, perspectives,} Laser Photonics Rev. \textbf{11}, 1600295 (2017).

\bibitem{KHORASANINEJAD_Science_metalenses_2017}
M.~Khorasaninejad and F.~Capasso, \enquote{Metalenses: Versatile
  multifunctional photonic components,} Science \textbf{358}, eaam8100 (2017).

\bibitem{MATEUS_IEEEPhotonicsTechnol.Lett._ultrabroadband_2004}
C.~F. Mateus, M.~C. Huang, Y.~Deng, A.~R. Neureuther, and C.~J. Chang-Hasnain,
  \enquote{Ultrabroadband mirror using low-index cladded subwavelength
  grating,} IEEE Photonics Technol. Lett. \textbf{16}, 518 (2004).

\bibitem{BRUCKNER_Phys.Rev.Lett._realization_2010}
F.~Br{\"u}ckner, D.~Friedrich, T.~Clausnitzer, M.~Britzger, O.~Burmeister,
  K.~Danzmann, E.-B. Kley, A.~T{\"u}nnermann, and R.~Schnabel,
  \enquote{Realization of a monolithic high-reflectivity cavity mirror from a
  single silicon crystal,} Phys. Rev. Lett. \textbf{104}, 163903 (2010).

\bibitem{FATTAL_NatPhoton_flat_2010}
D.~Fattal, J.~Li, Z.~Peng, M.~Fiorentino, and R.~G. Beausoleil, \enquote{Flat
  dielectric grating reflectors with focusing abilities,} Nat. Photonics
  \textbf{4}, 466 (2010).

\bibitem{YANG_NanoLett._dielectric_2014}
Y.~Yang, W.~Wang, P.~Moitra, I.~I. Kravchenko, D.~P. Briggs, and J.~Valentine,
  \enquote{Dielectric meta-reflectarray for broadband linear polarization
  conversion and optical vortex generation,} Nano Lett. \textbf{14}, 1394
  (2014).

\bibitem{ESFANDYARPOUR_Nat.Nanotechnol._metamaterial_2014}
M.~Esfandyarpour, E.~C. Garnett, Y.~Cui, M.~D. McGehee, and M.~L. Brongersma,
  \enquote{Metamaterial mirrors in optoelectronic devices,} Nat. Nanotechnol.
  \textbf{9}, 542 (2014).

\bibitem{ARBABI_Nat.Photonics_planar_2017}
A.~Arbabi, E.~Arbabi, Y.~Horie, S.~M. Kamali, and A.~Faraon, \enquote{Planar
  metasurface retroreflector,} Nat. Photonics \textbf{11}, 415 (2017).

\bibitem{ZHU_Optica_flexible_2015}
L.~Zhu, J.~Kapraun, J.~Ferrara, and C.~J. Chang-Hasnain, \enquote{Flexible
  photonic metastructures for tunable coloration,} Optica \textbf{2}, 255
  (2015).

\bibitem{LIN_Sci.Rep._optical_2017}
D.~Lin, M.~Melli, E.~Poliakov, P.~S. Hilaire, S.~Dhuey, C.~Peroz, S.~Cabrini,
  M.~Brongersma, and M.~Klug, \enquote{Optical metasurfaces for high angle
  steering at visible wavelengths,} Sci. Rep. \textbf{7}, 2286 (2017).

\bibitem{AYDIN_Nat.Commun._broadband_2011}
K.~Aydin, V.~E. Ferry, R.~M. Briggs, and H.~A. Atwater, \enquote{Broadband
  polarization-independent resonant light absorption using ultrathin plasmonic
  super absorbers,} Nat. Commun. \textbf{2}, 517 (2011).

\bibitem{brongersma2014light}
M.~L. Brongersma, Y.~Cui, and S.~Fan, \enquote{Light management for
  photovoltaics using high-index nanostructures,} Nat. Materials \textbf{13},
  451 (2014).

\bibitem{TAME_NatPhys_quantum_2013}
M.~S. Tame, K.~R. McEnery, {\c S}.~K. {\"O}zdemir, J.~Lee, S.~A. Maier, and
  M.~S. Kim, \enquote{Quantum plasmonics,} Nat. Phys. \textbf{9}, 329 (2013).

\bibitem{xia_twodimensional_2014}
F.~Xia, H.~Wang, D.~Xiao, M.~Dubey, and A.~Ramasubramaniam,
  \enquote{Two-dimensional material nanophotonics,} Nat. Photonics \textbf{8},
  899 (2014).

\bibitem{KHANIKAEV2017Nat.Photonics}
A.~B. Khanikaev and G.~Shvets, \enquote{Two-dimensional topological photonics,}
  Nat. Photonics \textbf{11}, 763 (2017).

\bibitem{LU_NatPhoton_topological_2014}
L.~Lu, J.~D. Joannopoulos, and M.~Solja{\v c}i{\'c}, \enquote{Topological
  photonics,} Nat. Photonics \textbf{8}, 821 (2014).

\bibitem{LI_Nat.Rev.Mater._nonlinear_2017}
G.~Li, S.~Zhang, and T.~Zentgraf, \enquote{Nonlinear photonic metasurfaces,}
  Nat. Rev. Mater. \textbf{2}, 17010 (2017).

\bibitem{MAGUID2017Science}
E.~Maguid, M.~Yannai, A.~Faerman, I.~Yulevich, V.~Kleiner, and E.~Hasman,
  \enquote{Disorder-induced optical transition from spin {{Hall}} to random
  {{Rashba}} effect,} Science \textbf{358}, 1411 (2017).

\bibitem{RYBIN_Nat.Commun._phase_2015}
M.~V. Rybin, D.~S. Filonov, K.~B. Samusev, P.~A. Belov, Y.~S. Kivshar, and
  M.~F. Limonov, \enquote{Phase diagram for the transition from photonic
  crystals to dielectric metamaterials,} Nat. Commun. \textbf{6}, 10102 (2015).

\bibitem{segev2013anderson}
M.~Segev, Y.~Silberberg, and D.~N. Christodoulides, \enquote{Anderson
  localization of light,} Nat. Photonics \textbf{7}, 197 (2013).

\bibitem{FENG2017Nat.Photonics}
L.~Feng, R.~El-Ganainy, and L.~Ge, \enquote{Non-{{Hermitian}} photonics based
  on parity\textendash{}time symmetry,} Nat. Photonics \textbf{11}, 752 (2017).

\bibitem{BARREDA_SciRep_light_2017}
{\'A}.~I. Barreda, Y.~Guti{\'e}rrez, J.~M. Sanz, F.~Gonz{\'a}lez, and
  F.~Moreno, \enquote{Light guiding and switching using eccentric core-shell
  geometries,} Sci. Rep. \textbf{7}, 11189 (2017).

\bibitem{ZHAI_Science_scalablemanufactured_2017-1}
Y.~Zhai, Y.~Ma, S.~N. David, D.~Zhao, R.~Lou, G.~Tan, R.~Yang, and X.~Yin,
  \enquote{Scalable-manufactured randomized glass-polymer hybrid metamaterial
  for daytime radiative cooling,} Science \textbf{355}, 1062 (2017).

\bibitem{JHA2015Phys.Rev.Lett.}
P.~K. Jha, X.~Ni, C.~Wu, Y.~Wang, and X.~Zhang, \enquote{Metasurface-enabled
  remote quantum interference,} Phys. Rev. Lett. \textbf{115}, 025501 (2015).

\bibitem{JHA2017ACSPhotonics}
P.~K. Jha, N.~Shitrit, J.~Kim, X.~Ren, Y.~Wang, and X.~Zhang,
  \enquote{Metasurface-mediated quantum entanglement,} ACS Photonics,
  doi:10.1021/acsphotonics.7b01241  (2017).

\bibitem{SMIRNOVA_Optica_multipolar_2016}
D.~Smirnova and Y.~S. Kivshar, \enquote{Multipolar nonlinear nanophotonics,}
  Optica \textbf{3}, 1241 (2016).

\bibitem{CUMMER2008Phys.Rev.Lett.}
S.~A. Cummer, B.-I. Popa, D.~Schurig, D.~R. Smith, J.~Pendry, M.~Rahm, and
  A.~Starr, \enquote{Scattering theory derivation of a {3D} acoustic cloaking
  shell,} Phys. Rev. Lett. \textbf{100}, 024301 (2008).

\bibitem{ZHANG2008Phys.Rev.Lett.}
S.~Zhang, D.~A. Genov, C.~Sun, and X.~Zhang, \enquote{Cloaking of matter
  waves,} Phys. Rev. Lett. \textbf{100}, 123002 (2008).

\bibitem{FARHAT2008Phys.Rev.Lett.}
M.~Farhat, S.~Enoch, S.~Guenneau, and A.~B. Movchan, \enquote{Broadband
  cylindrical acoustic cloak for linear surface waves in a fluid,} Phys. Rev.
  Lett. \textbf{101}, 134501 (2008).

\bibitem{LIAO2012Phys.Rev.Lett.}
B.~Liao, M.~Zebarjadi, K.~Esfarjani, and G.~Chen, \enquote{Cloaking core-shell
  nanoparticles from conducting electrons in solids,} Phys. Rev. Lett.
  \textbf{109}, 126806 (2012).

\bibitem{BRULE2014Phys.Rev.Lett.}
S.~Br{\^u}l{\'e}, H.~Javelaud, E.\, S.~Enoch, and S.~Guenneau,
  \enquote{Experiments on seismic metamaterials: molding surface waves,} Phys.
  Rev. Lett. \textbf{112}, 133901 (2014).

\bibitem{CLOS2016Phys.Rev.Lett.}
G.~Clos, D.~Porras, U.~Warring, and T.~Schaetz, \enquote{Time-resolved
  observation of thermalization in an isolated quantum system,} Phys. Rev.
  Lett. \textbf{117}, 170401 (2016).

\end{thebibliography}
\end{document}